\documentclass[acmsmall,screen]{acmart}

\AtBeginDocument{%
  }

\usepackage{tabularx}
\usepackage{multirow}
\usepackage{tablefootnote}
\usepackage{threeparttable}
\usepackage{pifont}
\usepackage[T1]{fontenc}
\usepackage{pbsi}
\usepackage{enumitem}
\usepackage[skins]{tcolorbox}
\usepackage[linesnumbered,ruled,vlined,noend]{algorithm2e}
\usepackage{caption}
\usepackage{listings}
\usepackage{tabularx}
\usepackage[normalem]{ulem}

\newcommand{\cmark}{\ding{51}}% check
\newcommand{\xmark}{\ding{55}}% x
\newcommand{\qmark}{\bsifamily ?}% ?

\newcommand{\toolname}{\textsc{PyRadar}}
\newcommand{\pysrc}{\textsc{py2src}}

\newcommand{\Code}[1]{\begin{small}\fontsize{9.5}{10}\selectfont\texttt{#1}\end{small}}

\makeatletter
\patchcmd\algocf@Vline{\vrule}{\vrule \kern-0.4pt}{}{}
\patchcmd\algocf@Vsline{\vrule}{\vrule \kern-0.4pt}{}{}
\makeatother

\newcommand{\FuncName}[1]{\normalfont\textsf{#1}}

\SetCommentSty{mycommfont}
\SetKwComment{Comment}{\ \#\ }{}
\SetKwProg{Func}{function}{}{end}

\tcbset{
  my box2/.style={
    enhanced,
    colframe=#1!80,
    colback=#1!5,
  },
}
\newtcolorbox{summary-rq}{
  my box2=black,
  boxrule=1pt,top=2pt,bottom=2pt,left=2pt,right=2pt
}

\begin{document}

\title[\toolname: Towards Automatically Retrieving and Validating PyPI Packages' Source Code Repository Information]{\toolname: Towards Automatically Retrieving and Validating Source Code Repository Information for PyPI Packages}

\author{Kai Gao}
\email{gaokai19@pku.edu.cn}
\orcid{0000-0002-0942-7890}
\affiliation{%
  \institution{School of Software \& Microelectronics, Peking University}
  \city{Beijing}
  \country{China}
}

\author{Weiwei Xu}
\email{xuww@stu.pku.edu.cn}
\orcid{0009-0004-3493-102X}
\author{Wenhao Yang}
\email{Yangwh@stu.pku.edu.cn}
\orcid{0000-0002-1005-1974}
\author{Minghui Zhou}
\authornote{Corresponding Author}
\email{zhmh@pku.edu.cn}
\orcid{0000-0001-6324-3964}
\affiliation{%
  \institution{School of Computer Science, Peking University}
  \city{Beijing}
  \country{China}
}

\begin{abstract}
A package's source code repository records the package's development history, which is critical for the use and risk monitoring of the package. 
However, a package release often misses its source code repository due to the separation of the package's development platform from its distribution platform. 
To establish the link, existing tools retrieve the release's repository information from its metadata, which suffers from two limitations: the metadata may not contain or contain wrong information. 
Our analysis shows that existing tools can only retrieve repository information for up to 70.5\% of PyPI releases. 
To address the limitations, this paper proposes \toolname, a novel framework that utilizes the metadata and source distribution to retrieve and validate the repository information for PyPI releases. 
We start with an empirical study to compare four existing tools on 4,227,425 PyPI releases and analyze phantom files (files appearing in the release's distribution but not in the release's repository) in 14,375 correct and 2,064 incorrect package-repository links. 
Based on the findings, we design \toolname~ with three components, i.e., Metadata-based Retriever, Source Code Repository Validator, and Source Code-based Retriever, that progressively retrieves correct source code repository information for PyPI releases. 
In particular, the Metadata-based Retriever combines best practices of existing tools and successfully retrieves repository information from the metadata for 72.1\% of PyPI releases. 
The Source Code Repository Validator applies common machine learning algorithms on six crafted features and achieves an AUC of up to 0.995. 
The Source Code-based Retriever queries World of Code with the SHA-1 hashes of all Python files in the release's source distribution and retrieves repository information for 90.2\% of packages in our dataset with an accuracy of 0.970. 
Both practitioners and researchers can employ the \toolname~ to better use PyPI packages.

\end{abstract}

\setcopyright{acmlicensed}
\acmJournal{PACMSE}
\acmYear{2024} \acmVolume{1} \acmNumber{FSE} \acmArticle{115} \acmMonth{7}\acmDOI{10.1145/3660822}




\maketitle

\section{Introduction}\label{s: intro}
In order to boost productivity and reduce cost, developers widely reuse ``the wheels''~\cite{Phuong2020-JSS, Larios2020-FSE, He2021-FSE}, i.e., reusing existing third-party packages rather than implementing the functionality from scratch. Many programming language (PL) communities provide centralized package registries (such as PyPI for Python and NPM for JavaScript) to facilitate the sharing and reuse of third-party packages, which host a substantially growing number of packages. As of September 2023, over 480 thousand Python packages have been published in PyPI.

Despite the benefits, reusing third-party packages also poses unique challenges to software development. First, given so many packages available, selecting the right one is laborious for developers~\cite{Larios2020-FSE}. Second, even if the right package is selected, how to estimate, monitor, and mitigate risks of the package, e.g., stop-of-maintenance~\cite{Valiev2018-FSE} and vulnerabilities~\cite{Decan2018-MSR, Pashchenko2018-ESEM, Markus2019-USENIX, Alfadel2021-SANER, Liu2022-ICSE, Pan2022-FSE}, is important yet challenging.
Merely using source code in the package's distribution is often insufficient in mitigating the above challenges. Practitioners and researchers commonly turn to the package's source code repository. On the one hand, development activity data recorded in the repository can help identify and mitigate some risks in the package.
For example, the number of stars is a critical factor when selecting third-party packages~\cite{Larios2020-FSE}; the number of commits, contributors, and issues are popular indicators of the package's maintenance state~\cite{Valiev2018-FSE}; researchers also mine undisclosed vulnerabilities from issues~\cite{Pan2022-FSE} and track patches for known vulnerabilities~\cite{Xu2022-FSE}.
On the other hand, the package's source code repository provides package users a place that the package registry lacks, allowing them to report bugs~\cite{Sebastiano2021-IST}, make contributions~\cite{Zhou2012-ICSE, Tsay2014-ICSE, Zhu2016-FSE}, and seek community help~\cite{Dabbish2012-CSCW, Zhou2015-TSE}.
Therefore, a package's source code repository not only complements the role of its distribution in estimating and mitigating risks of the package but also plays a crucial role in its usage. Locating the source code repository is vital for every package.

However, many major PL communities (Python, JavaScript, Java, etc.) follow the common practice of separating the code repository and distribution artifact of the package.
The separation offers numerous benefits, e.g. reducing the package's installation size~\cite{Ladisa2023-SP} and streamlining the build process~\cite{Wang2020-ICSE, Vu2021-FSE}, but disconnects the artifact from its source code repository.
The good news is, that developers usually use build tools to manage and publish distribution artifacts, and most build tools provide mechanisms empowering developers to declare the source code repository in the package metadata.
For example, \Code{setuptools}, the de facto build tool in the Python community, provides several optional keyword arguments (e.g., \Code{url}, \Code{project\_urls}) in the \Code{setup} function for package developers to declare package-related URLs in the package specification file such as \Code{setup.py}~\cite{PackagingUserGuide}. Then based on the package specification file, \Code{setuptools} automatically generates release metadata~\cite{Coremetadata} in the packaging process.
However, repository information in the metadata is still not the final answer to the package-repository linking problem.
First, since the package's source code repository information is not mandatory for build tools, \textit{package developers may not declare such information in the package specification file}. As shown in this paper, about 30\% of PyPI releases' metadata do not contain repository information.
Consequently, it is impossible to mine insights from source code repositories for these packages.
Second, \textit{package developers may declare wrong repository information in the package specification file} intentionally or unintentionally.
An extreme case is the prevalent typosquatting attack~\cite{Ohm2020-DIMVA, Ruian2021-NDSS, Ladisa2023-SP}.
Malicious package developers usually copy the metadata (including the repository information) of popular packages they masquerade as.
Another case is that developers did not change the placeholder repository URL.
\href{https://github.com/pypa/sampleproject}{gh:pypa/sampleproject} is a sample project that guides developers on packaging and distributing Python projects, but we find that 3,212 PyPI packages declare it as their source code repository.
Wrong source code repository information tends to mislead existing package monitoring tools such as Libraries.io~\cite{Librariesio} and open source insights~\cite{OpenSourceInsights}, and consequently, bias users' decisions.

Practitioners and researchers have implemented various tools ~\cite{Vu2021-ASE, Warehouse, Librariesio, ossgadget} to locate the repository for packages. However, these tools typically use the package metadata, which inevitably traps them into the two limitations mentioned above, i.e. wrong or unavailable repository information (that indicates where the repository is). To address the limitations, it is necessary to validate the repository information retrieved from the metadata and to utilize source code in the package to locate its repository when such information is unavailable in metadata.
To that end, we begin with a large-scale empirical study in PyPI and then leverage the discovered insights to design a tool to retrieve the correct repository location for PyPI releases. Specifically, the empirical study explores two research questions:

\begin{itemize}[leftmargin=*,topsep=0pt]
    \item \textbf{RQ1:} \textit{To what extent can existing tools retrieve source code repository information from the metadata and what are their differences?} Despite the plethora of metadata-based tools, there remains a lack of understanding about their capabilities. Thus, we propose this RQ to understand the status quo of techniques used to retrieve repository information from the metadata.
    \item \textbf{RQ2:} \textit{What are the phantom file differences between correct package-repository links and incorrect links?} Phantom files are files appearing in the release's distribution but not in the release's repository, possibly indicating whether a release is built from a repository. Assuming the package-repository link is correct, prior work~\cite{Vu2021-FSE} investigated phantom files and found that Python files were rarely phantom files in these links. However, to what extent phantom files differ between correct and incorrect package-repository links has not been investigated. Such understanding is vital for designing the tool that uses source code to validate and locate a package's repository.
\end{itemize}

We collect ecosystem-scale metadata of 4,227,425 PyPI releases to answer the questions.
We find that existing tools retrieve repository information from the metadata for up to 70.5\% of releases. We also identify several best practices such as URL redirection and searching from multiple information sources to locate the repository.
We propose a heuristic approach to collect 14,375 correct and 2,064 incorrect package-repository links, and a novel Git repository traversal algorithm to accurately identify phantom files. We find that the number of phantom files in incorrect links is significantly higher and incorrect links are more likely to contain phantom package specification files.

Inspired by the empirical findings, we propose \toolname, a novel framework that utilizes the release's metadata and source distribution to automatically \underline{r}etrieve \underline{a}nd vali\underline{da}te the source code \underline{r}epository information for PyPI releases. \toolname~ consists of three components: a \textbf{Metadata-based Retriever} similar to existing tools and two novel components to address the two limitations of existing tools, i.e., \textit{a \textbf{Source Code Repository Validator} to deal with the limitation of incorrect repository information in the metadata}, and \textit{a \textbf{Source Code-based Retriever} to address the limitation of missing repository information in the metadata}.
Specifically, the Metadata-based Retriever combines best practices of existing tools and retrieves repository information for 72.1\% of PyPI releases.
The Source Code Repository Validator validates the correctness of the repository information retrieved by the Metadata-based Retriever. It contributes six crafted features, two of which are derived from the findings of RQ2. Common machine learning algorithms can achieve an AUC of up to 0.995 on these features, demonstrating their effectiveness.
The Source Code-based Retriever uses source code in a release's source distribution to retrieve its repository from World of Code (WoC)~\cite{Ma2019-MSR, Ma2021-EMSE}, an infrastructure that collects almost all public Git repositories. Inspired by the findings of RQ2, we propose an efficient file hash-based repository retrieval algorithm. It retrieves repository information for 90.2\% of packages in our dataset with an accuracy of 0.970 and completes a retrieval in an average of 40 seconds, demonstrating its effectiveness and efficiency.

In summary, the major contributions of this paper are:
\begin{itemize}[leftmargin=*,topsep=0pt]
    \item We conduct the first large-scale empirical study to compare existing metadata-based tools and investigate phantom file differences between correct and incorrect package-repository links.
    \item We propose a heuristic approach to automatically and accurately collect correct and incorrect package-repository links to facilitate future research on this problem.
    \item We propose and evaluate \toolname, a novel framework that utilizes the metadata and source distribution to automatically retrieve and validate repository information for PyPI releases. It works for all PyPI releases (about 88\% of all PyPI releases) that provide source distributions.
\end{itemize}

\begin{figure}
    \centering
    \includegraphics[width=\linewidth]{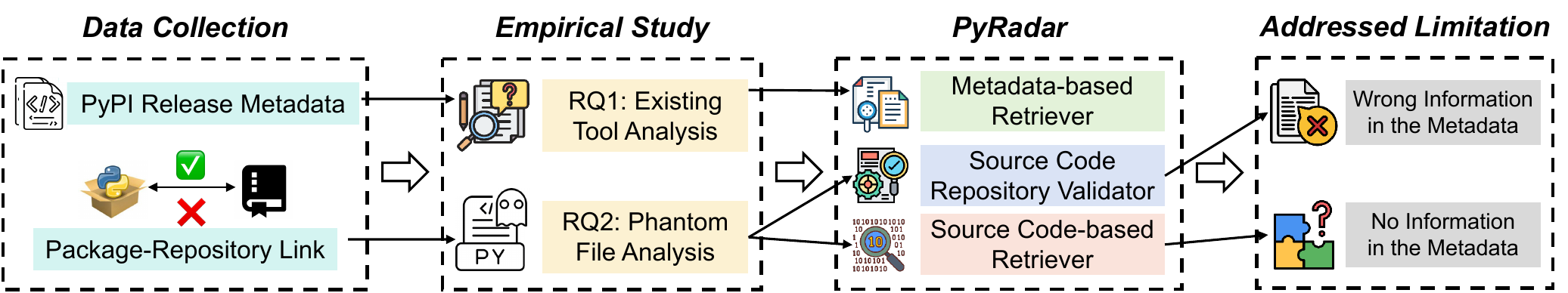}
    \vspace{-6mm}
    \caption{Overview of this study}
    \label{fig: overview}
    \vspace{-6mm}
\end{figure}

Figure~\ref{fig: overview} demonstrates the overview of this study, including the data collection (Section~\ref{s: dataset}), the empirical study (Section~\ref{s: empirical study}), and the \toolname~ framework addressing the two limitations (Section~\ref{s: tool}).

\section{Background and Related Work}

\subsection{Terminology}\label{ss: terminology}
Based on the official Python Packaging Glossary~\cite{Glossary} and prior work~\cite{Vu2021-FSE}, we define the following terminologies for the convenience of discussion.

\begin{itemize}[label={},itemindent=-2em,leftmargin=2em]
    \item \textit{Package} is a project registered on PyPI that `\textit{is intended to be packaged into a distribution}'. A package can be considered as a folder consisting of a collection of files, with package specification files at the top-level folder. A package has one or more releases.

    \item \textit{Release} is a snapshot of a package at a certain time point and is identified by a version identifier. A release consists of one or more distributions.

    \item \textit{Distribution} is a versioned archive file that contains the Python package. Distribution is what the user will download from the package registries and install. There are two types of distributions: \textit{source distribution} and \textit{built distribution}.

    \item \textit{Source distribution} is a distribution format that contains package specification files and all essential files needed to generate built distributions.

    \item \textit{Built distribution} is another distribution format containing only metadata and files that will be copied to the correct location on the user's system at installation time. It contains pre-compiled files such as \Code{.pyc} files and \Code{.so}/\Code{.dll} binary modules. However, Python files are not precompiled~\cite{Glossary}. Therefore, built distributions also contain Python source code.

    \item \textit{Package specification file} specifies build system information and release metadata. Currently, there exist multiple package specification files including \Code{pyproject.toml}, \Code{setup.py}, and \Code{setup.cfg}. Multiple package specification files can coexist in a package.

    \item \textit{Metadata} contains descriptive information of a package such as name, version, and relevant URLs.

\end{itemize}

\subsection{World of Code}\label{woc}
World of Code (WoC) is an infrastructure for mining version control system data across the entire open source software ecosystem.
It collects Git objects including commits, trees, and blobs~\cite{Gitobjects} from nearly exhaustive public Git repositories on dozens of code hosting platforms such as GitHub, Bitbucket, and GitLab.
Based on the collected Git objects, WoC provides several key-value databases for efficiently querying relationships between blobs, commits, repositories, and other relevant entities. For example, the blob-to-commit database maps a blob to all commits that introduced it; the commit-to-repository database maps a commit to all repositories that contain it. It is regularly updated and versioned. At the time of experimentation in this paper, the latest version of WoC was labeled as U and the data was collected in October 2021, containing over 173 million Git repositories, 3.1 billion commits, 12.5 billion trees, and 12.4 billion blobs~\cite{wocoverview}.

\subsection{Software Provenance}
Software provenance refers to the origin and history of a software artifact such as code snippets and files~\cite{Godfrey2015-SCP}.
In this sense, finding the package's source code repository is a kind of software provenance task.
Rousseau \textit{et al.}~\cite{Rousseau2020-EMSE} investigated the problem of file provenance in Software Heritage~\cite{Cosmo2017-iPRES}, an infrastructure for preserving software source code.
Hata \textit{et al.}~\cite{Hata2021-ICSE} studied how the same file evolves in different source code repositories.
Reid \textit{et al.}~\cite{Reid2022-ICSE} developed a tool VDiOS based on WoC to detect vulnerabilities induced by file reuse.
Wyss \textit{et al.}~\cite{Wyss2022-ICSE} proposed \textsc{unwrapper} to detect shrinkwrapped clones in NPM where a package duplicates or near-duplicates the code of another package.
However, all these works are conducted at file-level granularity, different from the distribution-level granularity (i.e., a collection of files) required by finding the package's repository.

Sun \textit{et al.}~\cite{Sun2023-EMSE} proposed an identifier-based approach to map Debian source packages written in Python to PyPI packages. They indexed identifiers (classes and method/function names) in all PyPI packages and found that 76\% of identifiers exist only in one package. Then they proposed an approach that randomly selects three non-frequent identifiers and queries the PyPI identifier corpus to locate the most probable PyPI packages. However, this work was conducted on the PyPI package corpus, which is different from and much smaller than the source code repository corpus.

\begin{table}[t]
\small
  \caption{Existing tools that retrieve the PyPI package release's source code repository information}
  \label{tab: existing tools}
  \renewcommand{\arraystretch}{1.1}
  \vspace{-3mm}
  \setlength{\tabcolsep}{4pt}
  \begin{tabular}{lcccc}
  \toprule
  Tool & Provider & Language & Platform & Open Source \\
  \midrule
  PyPI GitHub Statistics~\cite{Warehouse} & PyPI & Python & GitHub & \cmark \\
  OSSGadget OSS Find Source~\cite{OSSFindSource} & Microsoft & C\# & GitHub & \cmark \\
  Libraries.io~\cite{Librariesio} & Tidelift & Ruby & Multi-platform & \cmark \\
  \pysrc~\cite{Vu2021-ASE} & Duc-Ly Vu & Python & GitHub & \cmark \\
  Open Source Insights~\cite{OpenSourceInsights} & Google & \qmark & \qmark & \xmark \\
  Snyk Advisor~\cite{SnykAdvisor} & Snyk & \qmark & \qmark &  \xmark \\
  \bottomrule
  \end{tabular}
  \vspace{-6mm}
\end{table}

There are also some tools that automatically retrieve the package's source code repository information as summarized in Table~\ref{tab: existing tools}.
\begin{itemize}[leftmargin=*,topsep=0pt]
    \item PyPI GitHub Statistics~\cite{Warehouse} is provided by the PyPI website. Specifically, it detects GitHub repository URLs from the release's metadata and presents statistics for the retrieved GitHub repository such as the number of stars and forks in the sidebar of the package's PyPI page.
    \item OSSGadget~\cite{ossgadget} is a collection of software supply chain tools released by Microsoft, one of which is OSS Find Source which attempts to locate a release's source code repository on GitHub. Similar to PyPI GitHub Statistics, OSS Find Source retrieves the release's source code repository information from the metadata.
    \item Libraries.io~\cite{Librariesio} is an open source project maintained by Tidelift that collects package information from 32 package registries including PyPI. It also retrieves each package release's repository information from the metadata. Different from the above two tools, it detects repositories from multiple platforms.
    \item \pysrc~\cite{Vu2021-ASE} is a tool proposed by a researcher from the University of Trento. It retrieves GitHub repository URLs from multiple information sources, including the package metadata and the websites referenced by the metadata (i.e., the package's homepage and Readthedocs page). Then it returns the GitHub URL with the most occurrences.
    \item Open Source Insights~\cite{OpenSourceInsights} is a service developed and hosted by Google. It presents comprehensive information about the package such as vulnerabilities, dependencies, licenses, and the package's OpenSSF scorecard~\cite{OpenSSFScorecard} information based on the GitHub repository it detects. Since this tool is not open source~\cite{Frequent71:online}, we have little information about its implementation. But it claimed that the detected repository information \textit{is not guaranteed to be authoritative} as the package owner may list a link to any source code repository~\cite{BigQuery88:online}.
    \item Snyk Advisor~\cite{SnykAdvisor} is a service provided by the Snyk platform. It presents the package's maintenance and community information based on development activity data in the package's source code repository. It is also not open source. After manually inspecting several packages, we speculate this tool also retrieves repository information from the metadata and only extracts GitHub repository URLs.
\end{itemize}

To summarize, existing tools mainly retrieve the release's source code repository information from the metadata and do not validate the correctness of the retrieved repository information.

\section{Data Collection}\label{s: dataset}

We build two datasets to conduct this study: 1) the Metadata dataset for comparing existing metadata-based tools (RQ1) and providing necessary information for the rest part of this study; 2) the Package-Repository Link dataset, which is used to investigate phantom file differences between correct package-repository links and incorrect links (RQ2) and lay a foundation on the design and evaluation of the Source Code Repository Validator and the Source Code-based Retriever.

\textit{Metadata dataset}. We build this dataset with PyPI API~\cite{Warehouse} in March 2023. Specifically, we use the XML-RPC API to retrieve all packages registered on PyPI. We first get all its releases for each package and then get the metadata for each release using the JSON API. In total, we obtain metadata for 4,227,425 releases of 423,726 packages.

\textit{Package-Repository Link dataset}. To validate the retrieved repository information of a release is correct or not, we need to build a dataset consisting of correct package-repository links and incorrect links. However, since the repository information in the metadata may be incorrect, \textit{it is challenging to collect correct and incorrect package-repository links}. We, therefore, propose a heuristic approach to collect such data where we first collect correct package-repository links and then collect incorrect links based on the correct links.

\begin{figure}
\small
    \centering
    \includegraphics[width=0.9\linewidth]{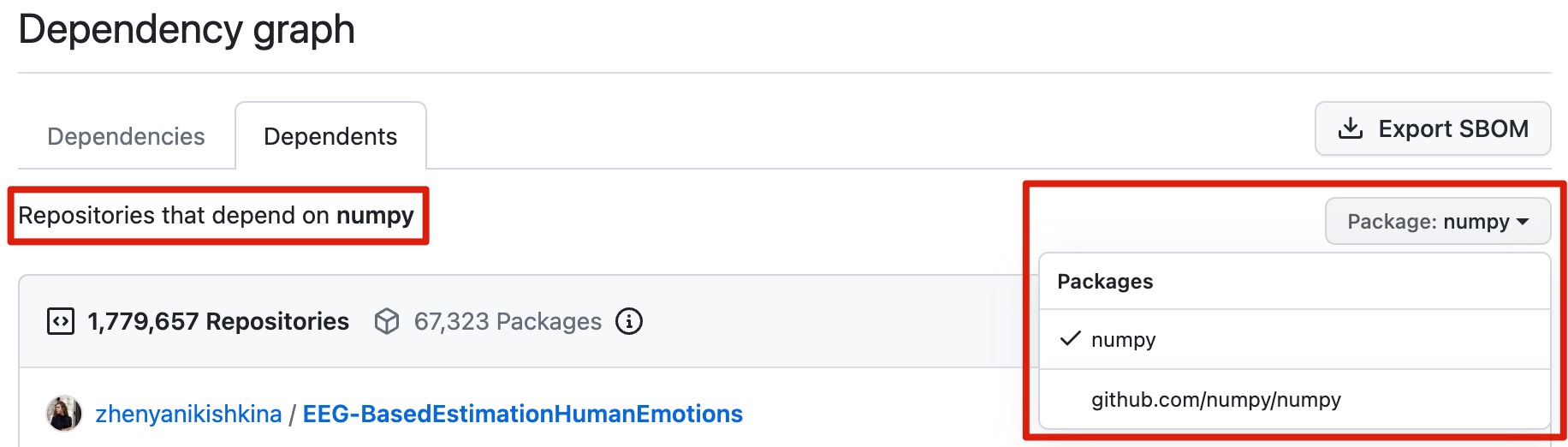}
    \vspace{-3mm}
    \caption{GitHub dependency graph page of the repository \href{https://github.com/numpy/numpy/network/dependents}{gh:numpy/numpy}}
    \label{fig:numpy}
    \vspace{-5mm}
\end{figure}

We turn to the GitHub dependency graph~\cite{GitHubDependencyGraph} (GDG for short) to collect correct links. GDG identifies the packages associated with a GitHub repository, as shown in Figure~\ref{fig:numpy}.
However, there are some problems when using GDG:
1) most GitHub repositories do not publish packages;
2) GDG detects packages across multiple packaging ecosystems, therefore packages published by different repositories may share the same name.
For example, \href{https://github.com/dmontagu/fastapi_client/network/dependents}{dmontagu/fastapi\_client} and \href{https://github.com/kevinastone/django-api-rest-and-angular/network/dependents}{kevinastone/django-api-rest-and-angular} both publish the \Code{example} package with the former as a Python package and the latter as a JavaScript package;
3) packages published by a repository may not be registered on PyPI and may even have the same name as PyPI packages. E.g., the \Code{teras} package published by \href{https://github.com/chantera/teras/network/dependents}{chantera/teras} collides with the \href{https://pypi.org/project/Teras/}{teras} package on PyPI.

To address these problems, we first collect all source code repositories with at least 100 stars and written in Python using GitHub search API~\cite{GitHubSearchAPI} (to tackle the first and second problem), since popular repositories are more likely to publish packages~\cite{Borges2016-ICSME, Wu2023-SANER}. We obtain 50,359 repositories in total.
Then we collect packages published by these repositories by crawling each repository's dependency graph page, resulting in 14,471 packages.
Next, we only keep packages registered on PyPI by aligning GitHub package names with those in the Metadata dataset (to tackle the third problem), leaving 12,463 packages published by 11,803 GitHub repositories.
We also add the top 4,000 most downloaded packages with their source code repository information retrieved by the Metadata-based Retriever to further enrich the dataset. We consider these packages' source code repository information correct since they have a high impact on the Python community and have been widely studied in prior work~\cite{Vu2021-FSE, Vu2021-ASE, Xu2023-ASE}. In total, we collect 14,375 correct links.

To evaluate the effectiveness of the approach to collecting correct links, we randomly sample 374 links (95\% confidence level and 5\% confidence interval).
For each link, we check:
1) if the package name is declared in any package specification file in the repository;
2) if the package's PyPI maintainer~\cite{PyPIMaintainer} (presented in the sidebar of the package's PyPI page) is a contributor to the repository (using common name abbreviation rules and user avatars);
3) if the repository or the documentation website referenced by the repository contains a link to the package's PyPI page or \Code{pip install} commands with the package name.
It is non-trivial to automate the three conditions accurately. Specifically, package developers can declare package names in the package specification file \Code{setup.py} in various ways, posing challenges in automating the first condition. The second condition, linking PyPI accounts and GitHub accounts, involves a trade-off between precision and recall due to incomplete and inconsistent information on different platforms~\cite{Vasilescu2013-ICSC,Silvestri2015-KDWeb,Fang2020-MSR}. The challenge of automating the third condition (searching for explicit mutual links between packages and repositories) is that relevant information is scattered and buried deep within repositories and related websites. Therefore, we choose manual checking over automated checking as it allows us to check the correct links with 100\% accuracy, which is important for verifying the validity of collected correct links.
We label a link as correct if it satisfies the first condition and one of the last two conditions. In total, 373, 369, and 349 links satisfy the three conditions respectively and 373 (99.7\%) links are labeled as correct. The remaining link is not labeled as correct due to the complex packaging process of the \Code{ansible} package~\cite{ansible}. It is noteworthy that the checking process is only a sufficient condition for a link to be correct. If a link does not meet the checking criteria, we can not assert it as incorrect. Overall, the manual inspection indicates the validity of the collected links.

Based on the correct links, we collect incorrect links as follows.
We presume that the maintainers of PyPI packages are developers in the package's source code repository responsible for publishing releases.
If two packages declare the same source code repository but have different maintainers, it is likely that one package's source code repository information is incorrect.
Under this assumption, we select packages that have the same source code repository as packages in the correct links but have different PyPI maintainers. In this way, we obtain 1,721 links. As noted in Section~\ref{s: intro}, over 3,000 packages declare their source code repository as \Code{gh:pypa/sampleproject} by mistake. Therefore, we expand the incorrect link data with these packages. Considering the large number of such packages, we randomly select 343 packages (95\% confidence level and 5\% confidence interval) to avoid them skewing the dataset. In total, we collect 2,064 incorrect package-repository links.

We verify the validity of incorrect links as follows. Since we are sure the incorrect correspondence between the 343 sampled packages and the \Code{gh:pypa/sampleproject} repository, we sample 314 links (95\% confidence level and 5\% confidence interval) from the rest of 1,721 links. For each link, we manually check if the package name is specified in the source code repository. If not, we label it as incorrect. In total, 312 (99.4\%) links are labeled as incorrect. The rest two links are correct due to the move of the package's source code repository~\cite{Edward2} and package renaming~\cite{coursera-dl}. Overall, the inspection results suggest that the collected incorrect links are valid.

\section{Empirical Study}\label{s: empirical study}

\subsection{RQ1: Existing Tool Analysis}\label{ss: rq1}
Given the many metadata-based tools available, a clear understanding of their effectiveness and discrepancies can help better retrieve the release's repository information from the metadata. Specifically, we aim to understand the number of PyPI releases for which existing tools can retrieve repository information and the differences in the retrieved repository information.

\subsubsection{Method}
We select PyPI GitHub Statistics, OSS Find Source, Libraries.io, and \pysrc~ for this RQ since they are open source, which enables us to make comparisons on the complete PyPI releases.
However, since these tools are tightly tied to their contexts and used in different ways, it is difficult to deploy them on our Metadata dataset. We, therefore, choose to carefully reimplement them to facilitate the evaluation of their capabilities on the metadata of 4,227,425 PyPI releases. To ensure that the reimplemented tools are consistent with the original ones, we test them with test cases from the original tools. Then, we deploy them on our dataset and obtain the repository information retrieved by these tools for each release.
To understand the discrepancies between these tools, we perform a stratified sampling of the differences in the retrieved repository information of the four tools and identify reasons for the differences based on the implementation of these tools. The number of sampled data for each pair of tools is shown in the parentheses of Table~\ref{tab: baseline diff}.

\begin{table}
\small
    \caption{The percentage of releases and packages for which the four reimplemented tools successfully retrieve repository information. Data in parentheses represents the adjusted percentage after URL redirection.}
  \label{tab: baseline results}
  \renewcommand{\arraystretch}{1.1}
  \vspace{-3mm}
  \begin{tabular}{lrrrr}
    \toprule
       & PyPI GitHub Statistics & OSS Find Source & Libraries.io & \pysrc \\
    \midrule
    Releases & 72.6\% (67.2\%) & 72.7\% (67.3\%) & 74.8\% (68.4\%) & 75.5\% (70.5\%) \\
    Packages & 68.4\% (60.9\%) & 68.5\% (61.0\%) & 70.6\% (62.2\%) & 70.2\% (63.1\%) \\
    \bottomrule
    \end{tabular}
    \vspace{-5mm}
\end{table}

\subsubsection{Results}
Table~\ref{tab: baseline results} presents the percentage of releases and packages for which the four reimplemented tools retrieve repository information.
It is worth noting that the retrieved repository information may be incorrect.
We can observe that existing tools can retrieve repository information for about 3/4 of releases and 70\% of packages, suggesting that substantial releases' metadata does not contain repository information. Libraries.io and \pysrc~ retrieve repository information for more releases since Libraries.io takes more code hosting platforms into account and \pysrc~ considers more information sources such as the homepage and Readthedocs page. The repository URLs retrieved by Libraries.io come from GitHub (3,032,984), GitLab (68,670), Bitbucket (53,668), SourceForge (5,918), ASF Subversion Server (2), and ASF GitBox Services (1). In the following, we only analyze the differences in the retrieved GitHub repository URLs.

The four tools retrieve different GitHub repository URLs for 511,480 (12.1\%) releases.
Table~\ref{tab: baseline diff} presents the percentage of releases for which the four tools retrieve different GitHub repository URLs.
The difference between PyPI GitHub Statistics and OSS Find Source is the least (0.16\%) since they only differ in how the GitHub repository URL is extracted. PyPI GitHub Statistics uses the URL scheme while OSS Find Source uses the regular expression.
\pysrc~ differs from the remaining three tools a lot (10.47\% $\sim$ 11.86\%).

\begin{table}
\small
    \caption{The percentage of releases for which the four reimplemented tools retrieve different GitHub repository URLs. Data in parentheses represents the number of sampled differences for each pair of tools.}
  \label{tab: baseline diff}
  \renewcommand{\arraystretch}{1.1}
  \vspace{-3mm}
  \begin{tabular}{lrrrr}
    \toprule
       & PyPI GitHub Statistics & OSS Find Source & Libraries.io & \pysrc \\
    \midrule
    PyPI GitHub Statistics & - & - & - & - \\
    OSS Find Source & 0.16\% (2) & - & - & - \\
    Libraries.io & 2.06\% (26) & 2.02\% (25) & - & - \\
    \pysrc & 10.47\% (104) & 10.43\% (103) & 11.86\% (121) & - \\
    \bottomrule
    \end{tabular}
    \vspace{-5mm}
\end{table}

Table~\ref{tab: diff reason} presents the seven identified reasons for the differences in the retrieved repository information of the four tools. URL redirection is the most common reason for the differences. Only \pysrc~ deals with URL redirection when retrieving repository URLs from some information sources.
We also present the adjusted percentage of releases and packages for which the four tools retrieve repository information after URL redirection in Table~\ref{tab: baseline results} (in parentheses). We can observe that existing tools can retrieve repository information for up to 70.5\% of releases and 63.1\% of packages. It suggests that URL redirection should be considered when retrieving repository information from the metadata due to the URL decay.
The strategy of searching \Code{project\_urls} field accounts for the second most differences. This field is an arbitrary map of names to URLs.
Libraries.io takes a rather conservative approach by searching only URLs whose names are in a predefined list, thus omitting repository URLs with other names.
The badge URL, Readthedocs page, and Homepage searched by \pysrc~account for about 1/4 of the differences.
Different URL extraction methods also lead to different retrieved repository information. Specifically, PyPI GitHub Statistics and \pysrc~ extract repository URLs according to the URL scheme while OSS Find Source and Libraries.io rely on regular expressions. We find that regular expressions are more robust since the URL information provided by developers may not strictly observe the URL scheme. It is noteworthy that these reasons may contribute to inaccuracies when retrieving repository information from the metadata. For example, when the metadata contains multiple URLs linking to repositories on code hosting platforms, these tools may select different URLs due to different retrieval strategies, leading to incorrect repository information retrieved by certain tools.

\begin{table}
\small
    \caption{Reasons for the differences in the retrieved repository information of the four reimplemented tools.}
  \label{tab: diff reason}
  \renewcommand{\arraystretch}{1.1}
  \setlength{\tabcolsep}{20pt}
  \vspace{-3mm}
  \begin{tabular}{lrr}
    \toprule
    Reason & Percentage \\
    \midrule
    URL redirection & 62.2\% (237) \\
    \Code{project\_urls} field searching & 12.6\% (48) \\
    Badge URL searching & 12.3\% (47) \\
    Readthedocs searching & 8.9\% (34) \\
    URL Extraction method  & 5.2\% (20) \\
    Homepage searching & 3.9\% (15) \\
    Other & 1.6\% (6) \\
    \bottomrule
    \end{tabular}
    \vspace{-5mm}
\end{table}

\begin{summary-rq}
\textbf{Answers for RQ1:}
The four reimplemented tools can retrieve repository information from the metadata for up to 70.5\% of releases and 63.1\% of packages. The percentage of differences in the repository information retrieved by the four tools ranges from 0.16\% to 11.86\%. Seven reasons induce the differences such as URL redirection and the \Code{project\_urls} searching strategy.

\vspace{-4pt}
\noindent\rule{\textwidth}{0.8pt}

\textbf{Implications:}
When retrieving repository information from the metadata, it is necessary to take the following elements into account: URL redirection, multiple code hosting platforms, multiple information sources that are contained in the metadata, and badges, homepage, and Readthedocs page that are referred by the metadata.
\end{summary-rq}

\subsection{RQ2: Phantom File Analysis}\label{ss: rq2}
Prior work has revealed that most Python files in the release distribution also exist in the release's repository~\cite{Vu2021-FSE}, which sheds light on how to validate the correctness of the release's repository information retrieved from the metadata and how to retrieve the release's repository using source code in the release's distribution.
Thus, our second research question aims to understand the differences in phantom files (files appearing in the release's distribution but not in the release's repository) between correct package-repository links and incorrect links.

\subsubsection{Method}
The basic idea of obtaining phantom files is to traverse all files in the release's distribution and the release's source code repository, calculate their hashes, and find the files whose hashes appear in the distribution but not in the repository.

As noted in Section~\ref{ss: terminology}, a release consists of one or more distributions and there are two kinds of distributions in a release: source distribution and built distribution.
We choose source distribution to obtain phantom files for two reasons:
\begin{itemize}[leftmargin=*,topsep=0pt]
    \item The number of releases providing source distributions is much higher than the number of releases providing built distributions (3,719,068 vs. 2,892,007). Therefore, the source distribution enables us to analyze more releases.
    \item For the 2,419,223 releases providing both source distributions and built distributions, we randomly select one release for each package, resulting in 243,518 releases. For each release, we compare files in the source distribution and built distribution. We find that files in the built distribution are all included in the source distribution for 216,741 (89.0\%) releases. Among the remaining releases, the built distributions mostly contain pre-compiled files (such as \Code{.pyc} files, \Code{.so} binary modules), which are usually not included in the repository.
\end{itemize}

It is straightforward to get hashes for all files in the distribution. We download the distribution, open it with the Python standard library \Code{tarfile} (for \Code{.tar.gz} distribution files) or \Code{zipfile} (for \Code{.zip} distribution files)~\cite{PEP527}, traverse files in it, and calculate each file's blob SHA-1 hash~\cite{Gitobjects}. We choose the blob SHA-1 hash because Git uses it to index files so that we don't have to calculate hashes for files in the repository.

However, \textit{it is challenging to traverse all files in the repository due to the complex Git-based development~\cite{Bird2009-MSR}, especially the adoption of Git submodules~\cite{GitSubmodules}}. Submodules, which are configured in the \Code{.gitmodules} file, allow developers to add a Git repository as a subfolder of another Git repository. We find submodules also popular in PyPI releases' repositories. Specifically, among the 3,047,112 releases for which the Metadata-based Retriever retrieves repository information, 159,360 (5.2\%) releases' repository use submodules such as \Code{numpy} and \Code{scipy}. Files in submodules are also packaged into distributions. If not considering submodules when traversing files in the repository, many false positive phantom files will be identified. To make things more complicated, submodules are updated over time. To properly deal with Git submodules and accurately identify phantom files, we propose a novel Git repository traversal algorithm (Algorithm~\ref{alg: repository-traversal}).

\begin{algorithm}
\small
  \DontPrintSemicolon
  \KwIn{A Git repository URL: $URL$}
  \KwOut{A set of tuples with the file name and the file's SHA-1 hash: $\mathcal{F}$}
  \Func{$\FuncName{traverse}(URL)$}{
    $\mathcal{F} \gets \emptyset$\;
    $repo\_dir \gets \FuncName{open\_repository}(URL)$\;
    $C \gets \FuncName{list\_commits}(repo\_dir, URL)$\;
    \For{commit $c \in C$} {
        $t \gets \FuncName{get\_root\_tree}(repo\_dir, c)$\;
        $\mathcal{F} \gets \mathcal{F} \cup \FuncName{traverse\_tree}(URL, t)$\;
    }
    \Return{$\mathcal{F}$}\;
  }
  \Func{$\FuncName{traverse\_tree}(URL, t)$}{
    $f \gets \emptyset$\;
    $repo\_dir \gets \FuncName{open\_repository}(URL)$\;
    $submodules \gets \emptyset$\;
    \For{entry $e \in \FuncName{list\_tree\_entries}(repo\_dir, t)$}{
        \If{$e$ is a blob object}{
            $f.add((e.filename, e.sha1))$\;
            \If{$e.filename = \Code{".gitmodules"}$}{
            $submodules \gets \FuncName{parse\_gitmodules}(e.content)$\;
            }
        }\ElseIf{$e$ is a tree object}{
            $f \gets f \cup \FuncName{traverse\_tree}(URL, e)$\;
        }\ElseIf{$e$ is a commit object}{
            $submodule\_url \gets submodules.get(e.path)$\;
            $submodule\_dir \gets \FuncName{open\_repository}(submodule\_url)$\;
            $submodule\_tree \gets \FuncName{get\_root\_tree}(submodule\_dir, e.sha1)$\;
            $f \gets f \cup \FuncName{traverse\_tree}(submodule\_url, submodule\_tree)$\;
        }
    }
    \Return{$f$}\;
  }
  \caption{Traversing files in a Git repository}
  \label{alg: repository-traversal}
\end{algorithm}

\begin{table}
\small
  \caption{Comparison of our algorithm with \textsc{LastPyMile}. $\bigstar$ marks repositories using submodules. }
  \label{tab: LastPyMile comparision}
  \renewcommand{\arraystretch}{1.2}
  \setlength{\tabcolsep}{4pt}
  \vspace{-3mm}
  \begin{tabular}{llrrrrr}
    \toprule
    \multirow{2}{*}{Package} & \multirow{2}{*}{Repository} & \multirow{2}{*}{\# of Distribution Files} & \multicolumn{2}{c}{\# of Repository Files} & \multicolumn{2}{c}{\# of Phantom Files}\\
    & & & \textsc{LastPyMile} & Ours & \textsc{LastPyMile} & Ours \\
    \midrule
    six & \href{https://github.com/benjaminp/six}{gh:benjaminp/six} & 11 & 743 & 743 & 1 & 1\\
    certifi & \href{https://github.com/certifi/python-certifi}{gh:certifi/python-certifi} & 10 & 240 & 240 & 1 & 1 \\
    numpy & $\bigstar$ \href{https://github.com/numpy/numpy}{gh:numpy/numpy} & 2,231 & 64,506 & 63,719 & 296 & 4\\
    scipy & $\bigstar$ \href{https://github.com/scipy/scipy}{gh:scipy/scipy} & 18,855 & 66,600 & 89,939 & 15,581 & 0\\
    \bottomrule
    \end{tabular}
    \vspace{-5mm}
\end{table}

This algorithm traverses all commits to obtain a complete list of files in the repository. We use the Git command: \Code{git cat-file -{}-batch-check -{}-batch-all-objects -{}-unordered} (line 4) to ensure that all commits are traversed. For each commit, we obtain the tree object it points (line 6) and traverse the tree in a recursive way (\FuncName{traverse\_tree}) to obtain all files in the commit snapshot (line 7). Specifically, we first list all entries in the tree object, where each entry consists of the SHA-1 hash of a blob, tree, or commit object with its associated mode, type, and filename. We process each entry as follows:
\begin{itemize}[leftmargin=*,topsep=0pt]
    \item If the entry points to a blob object, we simply record its name and SHA-1 hash (line 14-15). If the blob's filename is \Code{".gitmodules"}, we parse the path and URL of submodules configured in this file with the Python standard library \Code{configparser} (line 16-17).
    \item If the entry points to a tree object, we traverse files in it and merge the result (line 18-19).
    \item If the entry points to a commit object, which indicates a submodule, we get the submodule's Git URL, obtain the tree object pointed by the commit object, and traverse files in the tree object of the submodule (line 20-24).
\end{itemize}

We implement this algorithm with as many operations provided by Git as possible to ensure a correct implementation. We make a simple comparison with \textsc{LastPyMile} proposed in prior work~\cite{Vu2021-FSE}, which did not consider submodules when traversing the repository, on four popular packages, two of which do not use submodules in their repositories and the other two of which use submodules in their repositories. The comparison results (Table~\ref{tab: LastPyMile comparision}) indicate that our algorithm can properly deal with submodules and identify more accurate phantom files.

We use the Package-Repository Link dataset to conduct this RQ. Specifically, for each (correct and incorrect) link in the dataset, we first download the source distribution of the package's latest release and obtain the blob SHA-1 hashes of files in the distribution. Then we clone the source code repository and obtain all blob SHA-1 hashes following Algorithm~\ref{alg: repository-traversal}. Finally, we get phantom files by comparing SHA-1 hashes in the distribution and the repository.

\subsubsection{Results}
\begin{figure}
    \centering
    \includegraphics[width=0.95\linewidth]{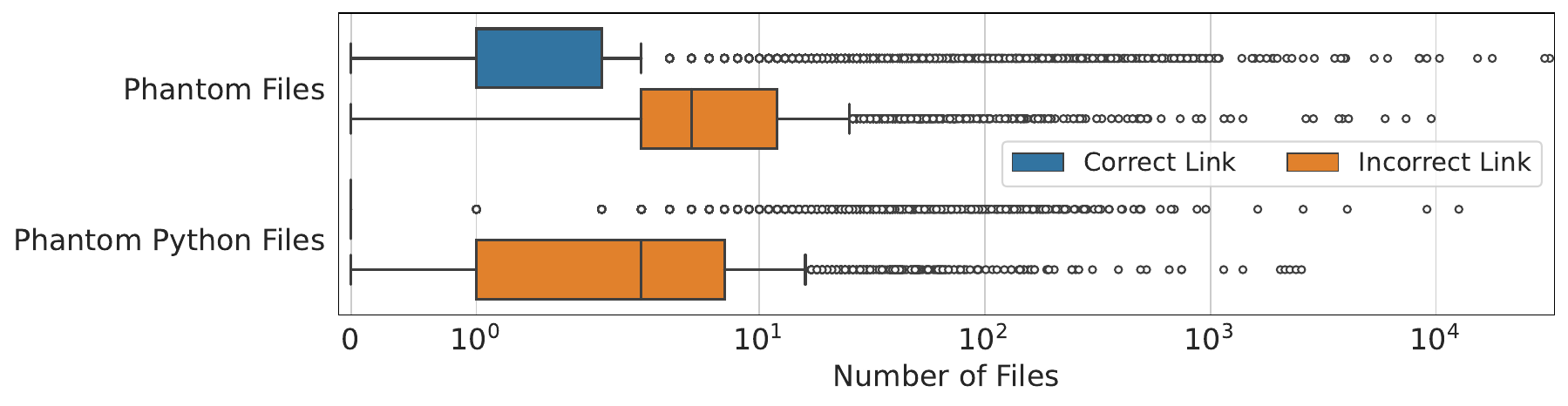}
    \vspace{-3mm}
    \caption{Distribution of the number of phantom files and phantom Python files in correct and incorrect links.}
    \label{fig: phantom file comparison}
    \vspace{-3mm}
\end{figure}

Figure~\ref{fig: phantom file comparison} demonstrates the distribution of the number of phantom files and phantom Python files in the correct and incorrect links. It can be clearly observed that the number of phantom files and phantom Python files in the correct links is much lower than that in the incorrect links, with the significance confirmed by the Mann-Whitney U test~\cite{mannwhitneyu}. Specifically, the median number of phantom files in the correct and incorrect links is 1 and 5 respectively, and the median number of phantom Python files is 0 and 3 respectively. It is also noteworthy that 75.3\% of the correct links do not have phantom Python files whereas 96.2\% of the incorrect links have at least one phantom Python file, suggesting that the number of phantom Python files may be useful to validate the correctness of a release's repository information and Python file in the release's source distribution may be used to retrieve the release's repository from WoC.

\begin{table}
\small
  \caption{The 10 most common files and Python files in source distributions of in correct package-repository links. The two numbers in the cell correspond to the file's inclusion rate and phantom rate, respectively.}
  \vspace{-3mm}
  \label{tab: phantom files}
  \renewcommand{\arraystretch}{1.2}
  \begin{tabular}{lrr|lrr}
  \toprule
  \textsc{Files} & \textsc{Correct} & \textsc{Incorrect} & \textsc{Python Files} & \textsc{Correct} & \textsc{Incorrect} \\
  \midrule
  \_\_init\_\_.py & 93.4\%, 6.8\% & 93.7\%, 48.5\% & \_\_init\_\_.py & 93.4\%, 6.8\% & 93.7\%, 48.5\% \\
  setup.cfg & 87.8\%, 98.1\% & 93.0\%, 99.3\% & setup.py & 94.7\%, 16.6\% & 93.0\%, 96.4\% \\
  setup.py & 94.7\%, 16.6\% & 93.0\%, 96.4\% & utils.py & 28.8\%, 4.1\% & 25.9\%, 31.1\% \\
  README.md & 58.5\%, 7.9\% & 55.5\%, 58.0\% & conf.py & 12.8\%, 4.4\% & 15.9\%, 54.1\% \\
  MANIFEST.IN & 54.7\%, 6.1\% & 49.1\%, 21.0\% & exceptions.py & 14.0\%, 4.5\% & 15.3\%, 24.1\% \\
  LICENSE & 54.3\%, 4.5\% & 43.5\%, 31.0\% & base.py & 13.6\%, 4.1\% & 12.8\%, 37.9\% \\
  README.rst & 33.5\%, 6.6\% & 37.2\%, 41.0\% & \_\_main\_\_.py & 13.7\%, 8.2\% & 12.4\%, 36.7\% \\
  utils.py & 28.8\%, 4.1\% & 25.9\%, 31.1\% & conftest.py & 9.0\%, 3.5\% & 12.1\%, 57.0\% \\
  pyproject.toml & 24.3\%, 9.2\% & 25.2\%, 64.6\% & models.py & 9.0\%, 3.2\% & 10.6\%, 37.6\% \\
  requirements.txt & 16.3\%, 5.2\% & 17.9\%, 40.9\% & version.py & 11.7\%, 18.6\% & 9.6\%, 51.0\% \\
  \bottomrule
  \end{tabular}
  \vspace{-3mm}
\end{table}

Table~\ref{tab: phantom files} presents the 10 most common files and Python files in source distributions of incorrect links. To derive effective features for identifying as much incorrect repository information for the release as possible, we calculate the inclusion rates and phantom rates of the files. The inclusion rate of a file is defined as the proportion of correct (incorrect) links that include the file in the source distribution compared to the total number of correct (incorrect) links. The phantom rate of a file is defined as the proportion of correct (incorrect) links in which the file is a phantom file compared to the number of correct (incorrect) links that include the file in the source distribution. Almost every correct link and incorrect link include the package specification file \Code{setup.py} or \Code{pyproject.toml} in the source distribution, as revealed by the inclusion rates in correct links (99.5\%) and incorrect links (99.2\%). However, the phantom rate of \Code{setup.py} or \Code{pyproject.toml} differs greatly between correct (16.6\%) and incorrect (97.0\%) links. The results indicate that the presence of a phantom package specification file (\Code{setup.py} or \Code{pyproject.toml}) could serve as an effective feature in identifying incorrect repository information for the release: an inclusion rate close to 1 indicates that it is calculable for almost every package-repository link and the high phantom rate in incorrect links indicates that it achieves a high recall in identifying incorrect repository information. Despite the high phantom rate of README (\Code{README.md} or \Code{README.rst}) or LICENSE in incorrect links compared to correct links, the presence of a phantom README or LICENSE is not effective in identifying incorrect repository information for two reasons: 1) the relatively low inclusion rates (91.0\% in correct links and 91.9\% in incorrect links) suggest that it is incalculable for nearly 10\% of links; 2) the low phantom rate (51.6\%) in incorrect links indicates its low recall.
Notably, the phantom rates of \Code{setup.cfg} (another package specification file) in correct and incorrect links are both close to 1. We manually inspect 10 correct links with the phantom \Code{setup.cfg} file and find that 7 of them do not contain the \Code{setup.cfg} file in the repository, indicating that most of the phantom \Code{setup.cfg} files are generated in the build process.
Overall, the results indicate that whether the package specification file \Code{setup.py} or \Code{pyproject.toml} is a phantom file may be used to identify incorrect repository information for the release.

\begin{summary-rq}
\textbf{Answers for RQ2:}
The number of phantom files and phantom Python files in the incorrect links is significantly higher than in the correct links. Python files are generally not phantom files in the correct links. The percentage of phantom package specification files (\Code{setup.py} or \Code{pyproject.toml}) in the incorrect links is much higher than in the correct links.

\vspace{-4pt}
\noindent\rule{\textwidth}{0.8pt}

\textbf{Implications:}
The number of phantom Python files and whether the package specification file \Code{setup.py} or \Code{pyproject.toml} is a phantom file may be useful features to validate the correctness of a release's repository information. Retrieving a release's repository from WoC using only Python files in the release's source distribution may be sufficient and efficient.
\end{summary-rq}

\section{The \toolname~ Approach}\label{s: tool}

\begin{figure}
    \centering
    \includegraphics[width=0.95\linewidth]{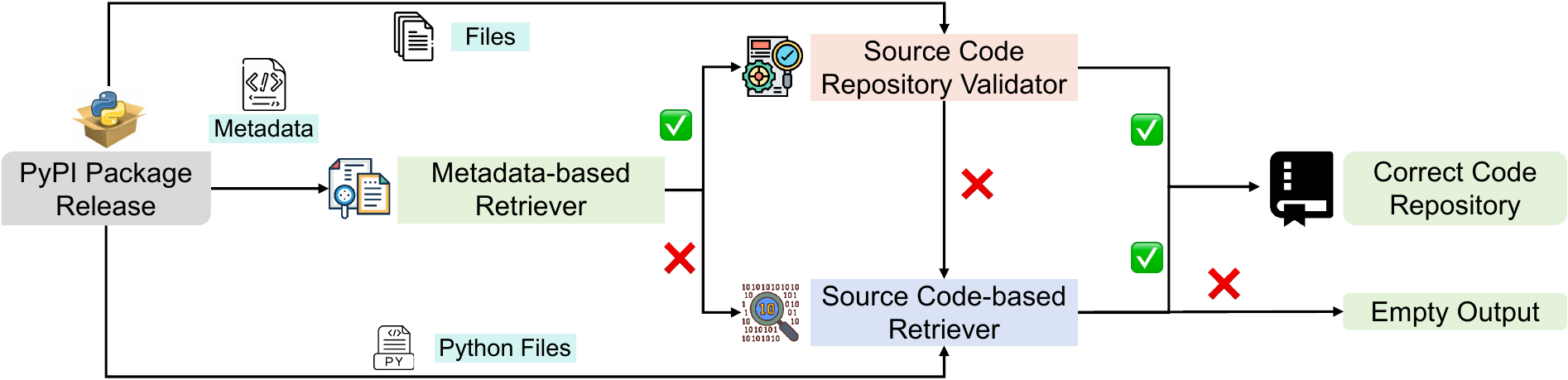}
    \vspace{-3mm}
    \caption{Overview of the \toolname~ framework}
    \label{fig: framework}
    \vspace{-4mm}
\end{figure}

Inspired by the empirical findings from Section~\ref{s: empirical study}, we propose \toolname, a framework that utilizes the release's metadata and source code to automatically retrieve and validate the release's source code repository information. As shown in Figure~\ref{fig: framework}, \toolname~ consists of three components:
\begin{itemize}[leftmargin=*,topsep=0pt]
    \item Metadata-based Retriever. It retrieves repository information from the release's metadata.
    \item Source Code Repository Validator. It validates the correctness of the release's repository information retrieved by the Metadata-based Retriever. If the repository information is validated as correct, it will then be output.
    \item Source Code-based Retriever. If the Metadata-based Retriever fails to retrieve repository information from the release metadata or the Source Code Repository Validator concludes that the release's repository information retrieved from the metadata is incorrect, this component retrieves the release's repository from WoC using files in the release's source distribution. If this component fails to retrieve a repository, the output is empty.
\end{itemize}

In this section, we elaborate on the design and evaluation of each component.

\subsection{Metadata-based Retriever}
\subsubsection{Design}
We design this component following the best practices learned from existing metadata-based tools (Section~\ref{ss: rq1}).
It first searches for repository URLs with regular expressions from the \Code{url}, \Code{download\_url}, and \Code{project\_urls} fields in the release metadata sequentially. If any repository URL is found, it returns the redirected repository URL.
Otherwise, it searches the \Code{description} field in the release metadata for repository URLs and badge URLs. It only returns the redirected URL whose repository name is exactly the same as the package name after removing non-alphanumeric characters.
If still no repository URL is found, it searches for the homepage URL and documentation page URL in the \Code{project\_urls} field, scrapes the homepage and documentation page, extracts repository URLs in them, and returns the redirected URL with the repository name exactly the same as the package name after removing non-alphanumeric characters.
In the current implementation, it considers repositories on the three most popular code hosting platforms, i.e., GitHub, GitLab, and Bitbucket.
It resolves URL redirections for GitHub repositories via the GitHub Repository API and for Bitbucket and GitLab repositories via HTTP requests.

\subsubsection{Evaluation}
We implement this component in Python and run it on the entire Metadata dataset. It successfully retrieves source code repository information for 72.1\% of the releases (79.1\% before URL redirection), 1.6\% higher than the existing state-of-the-art tool \pysrc. The improvement of the Metadata-based Retriever over existing tools is minor, possibly because only about 70\% of the releases' metadata contain repository information. Similar to other metadata-based tools, our Metadata-based Retriever will retrieve incorrect links when 1) the metadata does not contain a correct repository URL, or 2) the metadata contains a correct repository URL but the retriever selects another URL. Therefore, we design the Source Code Repository Validator to validate the correctness of the repository information retrieved by the Metadata-based Retriever.

\subsection{Source Code Repository Validator}
\subsubsection{Design}
The goal of this component is to validate the correctness of the repository information retrieved from the metadata. We design it as a classifier where the input is a pair of a release and a repository and the output is the probability of the input being an incorrect link.
Inspired by the findings of RQ2 (Section~\ref{ss: rq1}) and prior work~\cite{Taylor2020-NSS, Vu2020-EuroS, Vu2021-ASE}, we derive six crafted features (Table~\ref{tab: features}).
\begin{itemize}[leftmargin=*,topsep=0pt]
    \item According to Section~\ref{ss: rq2}, the correct link and incorrect link differ a lot on the number of phantom Python files and whether the package specification file \Code{setup.py} or \Code{pyproject.toml} changes. Therefore, we propose the \Code{\#phantom\_pyfiles} and \Code{pkg\_spec\_change} feature.
    \item Prior work~\cite{Vu2021-ASE} manually checked the alignment of the repository tags and PyPI releases' version identifiers to determine a reliable repository for a release. After manually inspecting the release's version identifier and the repository's tag in some correct links, we find that the repository tags usually end with the release's version identifier (e.g., v1.2.1 and 1.2.1). Therefore, we propose the \Code{tag\_alignment} feature which is set to 1 if any tag in the repository ends with the release's version identifier otherwise 0.
    \item Name similarity is also used to determine a release's reliable repository~\cite{Vu2021-ASE} and identify typosquatting packages~\cite{Taylor2020-NSS, Vu2020-EuroS}. Therefore, we consider the \Code{name\_similarity} feature useful for validating the correctness of the release's repository information as well. This feature is measured by the normalized Levenshtein similarity~\cite{levenshtein1966binary} between the package name and the repository name, ranging from $[0, 1]$.
    \item We also take the release's maintainer information into account, i.e., the \Code{\#maintainers} and \Code{\#maintainer\_pkgs} feature. Intuitively, if a package is maintained by more experienced developers, it is less likely to declare incorrect repository information.
\end{itemize}

\begin{table}
\small
    \caption{Description of the six crafted features.}
    \vspace{-3mm}
  \label{tab: features}
  \renewcommand{\arraystretch}{1.2}
  \begin{tabularx}{\linewidth}{m{0.2\linewidth}m{0.7\linewidth}}
    \toprule
    Feature & Description \\
    \midrule
    \#phantom\_pyfiles & The number of phantom Python files\\
    pkg\_spec\_change & Whether the package specification file \Code{setup.py} or \Code{pyproject.toml} is a phantom file or not\\
    tag\_alignment & Whether the release's version identifier aligns with any tag in the repository.\\
    name\_similarity & The normalized Levenshtein similarity between the package name and the repository name, ranging from $[0, 1]$.\\
    \#maintainers & The number of maintainers of the package \\
    \#maintainer\_pkgs & The number of packages maintained by the package's maintainers \\
    \bottomrule
    \end{tabularx}
    \vspace{-3mm}
\end{table}

We collect the six features for each link in the Package-Repository Link dataset,\footnote{We only select the latest release for each package.} including 14,375 correct links (labeled as 0) and 2,064 incorrect links (labeled as 1). Then we train machine learning models on the collected features. Specifically, we try seven commonly used models including Logistic Regression, SVM, Decision Tree, Random Forest, AdaBoost~\cite{freund1997decision}, Gradient Boosting Decision Tree~\cite{friedman2001greedy}, and XGBoost~\cite{DBLP:conf/kdd/ChenG16}. Due to the imbalanced distribution of correct and incorrect links in the dataset, we employ resampling techniques on the training set for all models, which is a common practice to address the issue of unbalanced samples~\cite{Tian2022-ICSE, DBLP:conf/icse/XiaoHXTDZ22}. We tune hyperparameters for each model with grid searching and select the best-performing model. Since the dataset is imbalanced and the relative ranking of incorrect links matters more, we choose AUC (Area under the ROC Curve)~\cite{hanley1982use} as the performance metric. AUC provides an evaluation of the model's performance across various classification thresholds and quantifies the probability that a random positive sample will have a higher ranking than a random negative sample~\cite{Melo2013}.

\subsubsection{Evaluation}\label{s5.2.2}
Table~\ref{tab: Validator performance} presents the optimal performance of each model in identifying the incorrect link. We also present the accuracy, precision, and recall at a threshold of 0.5 as a reference.
From Table~\ref{tab: Validator performance}, we can observe that either model archives an AUC higher than 0.95, possibly indicating the effectiveness of the six features. Random Forest outperforms the other models with an AUC of 0.995. It also achieves the highest recall score (0.989), suggesting that it can discover the most incorrect links in the test set. So we choose it as the classification model of this component. We also perform feature importance analysis to demonstrate how important each feature is for the fitted Random Forest model. Specifically, we use the permutation importance~\cite{Breiman2001-ML}, a well-known technique to measure the contribution of each feature to the fitted model. It is calculated by randomly shuffling the values of a feature and observing the resulting degradation of the model’s score. The results show that \Code{name\_similarity} is the most predictive feature (importance value: 0.117) followed by \Code{pkg\_spec\_change} (importance value: 0.028) and \Code{\#phantom\_pyfiles} (importance value: 0.026). The importance scores for the remaining three features are all below 0.010.

\begin{table}
\small
    \caption{Performance of the seven machine learning models. Accuracy, precision, and recall are calculated at a threshold of 0.5.}
    \vspace{-3mm}
  \label{tab: Validator performance}
  \renewcommand{\arraystretch}{1.2}
  \begin{tabular}{lrrrr}
    \toprule
    Approaches & AUC & Accuracy & Precision & Recall \\
    \midrule
    Logistic Regression & 0.963 & 0.890 & 0.744 & 0.871 \\
    SVM & 0.981 & 0.928 & 0.814 & 0.933 \\
    Decision Tree & 0.982 & 0.966 & 0.918 & 0.954 \\
    Random Forest & \textbf{0.995} & 0.973 & 0.913 & \textbf{0.989} \\
    AdaBoost & 0.992 & 0.950 & 0.846 & 0.984 \\
    Gradient Boosting Decision Tree & 0.992 & \textbf{0.974} & 0.936 & 0.963 \\
    XGBoost & 0.991 & 0.972 & \textbf{0.937} & 0.956 \\
    \bottomrule
    \end{tabular}
    \vspace{-3mm}
\end{table}

We then use this component to validate the repository information retrieved by the Metadata-based Retriever. The Metadata-based Retriever successfully retrieves repository information for 3,047,112 releases (273,273 packages). We exclude packages in the Package-Repository Link dataset and select the latest release that provides the source distribution for each package, resulting in 228,448 releases.
We then use the trained Random Forest model to output the probability of the link between the release and the retrieved repository being an incorrect link.
We manually inspect the top 100 links with the highest probability and find that 85 (85\%) of these links are indeed incorrect. For the remaining 15 correct links, we find the package name differs a lot from the repository name ($name\_similarity < 0.25$), which may lead to a high probability.

\subsection{Source Code-based Retriever}
\subsubsection{Design}
This component relies on the World of Code (WoC)~\cite{Ma2019-MSR, Ma2021-EMSE} infrastructure due to its extensive collection of public repositories and convenient APIs. However, \textit{the millions of repositories in WoC pose a great challenge in efficiently locating a release's correct repository based on source code}. To tackle this challenge, we propose a simple and efficient file hash-based algorithm based on the findings of RQ2. Specifically, this component retrieves a release's repository from WoC using Python files in its source distribution. We only use Python files since most of them are present in both the release's source distribution and the release's repository (Section~\ref{ss: rq2}), thus establishing a good link between the release and the repository. We use correct links in the Package-Repository Link dataset to design and evaluate this component. Among the 14,375 correct links, the repositories of 12,375 (86.1\%) links are indexed by WoC.

This component first retrieves candidate repositories from WoC following the \FuncName{get\_candidate} function in Algorithm~\ref{alg: woc-retriever}. For each Python file in the release's source distribution, it gets the first commit that introduces this file via the blob-to-commit database (line 4-5). Then it gets all repositories containing this commit (thus the file) via the commit-to-repository database (line 6).
To speed up the retrieval and reduce the candidate set, we only consider files that do not exist in many repositories controlled by a threshold $blob\_uniqueness$ (line 7-8). Finally, we rank candidate repositories by the number of Python files in the source distribution that they contain.

\begin{algorithm}
\small
  \DontPrintSemicolon
  \KwIn{A release's source distribution: $sdist$}
  \KwOut{The most probable repository: $r$}
  \Func{$\FuncName{get\_candidate}(sdist, blob\_uniqueness)$}{
    $\mathcal{R} \gets \emptyset$\;
    \For{Python file $f \in sdist$} {
        $blob\_sha \gets \FuncName{calculate\_sha}(f)$\;
        $c \gets \FuncName{get\_first\_commit}(blob\_sha)$\;
        $repos \gets \FuncName{query\_c2p}(c)$\;
        \If{len(repos) $\le$ blob\_uniqueness}{
            % $repos \gets repos \cup \FuncName{query\_p2P}(repos)$\;
            $\mathcal{R} \gets \mathcal{R} \cup \langle repos, f \rangle$\;
        }
    }
    $R.rank()$\;
    \Return{$\mathcal{R}$}\;
  }
  \Func{$\FuncName{get\_most\_probable}(sdist, blob\_uniqueness, topn, name\_similarity)$}{
  $repos \gets \emptyset$\;
  $\mathcal{R} \gets \FuncName{get\_candidate}(sdist, blob\_uniqueness)$\;
  \For{repo $r \in \FuncName{select\_topn}(\mathcal{R}, topn)$}{
    $repos \gets repos \cup \FuncName{defork}(r)$\;
  }
  $r \gets \FuncName{most\_common}(repos)$\;
  \If{$\FuncName{similarity}(r, sdist.name) < name\_similarity$}{
    $r \gets null$\;
  }
  \Return{$r$}\;
  }
  \caption{Retrieving repository from WoC}
  \label{alg: woc-retriever}
\end{algorithm}

To select the correct candidate, we manually inspect 373 links (95\% confidence level and 5\% confidence interval) from the 12,375 links and find that the top-ranked candidates are either the correct repository or forks of the correct repository in most (360, 96.5\%) links. Therefore, we choose the $topn$ candidate repositories, find their upstream forked repository, and select the most common repository.
To ensure the correctness of the retrieved repository, we only return the repository whose name similarity with the release's package name is above a threshold $name\_similarity$.
Increasing the threshold improves the accuracy but reduces the percentage of releases for which this component can retrieve repositories from WoC (i.e., the coverage), while decreasing it increases the coverage but reduces the accuracy.
We heuristically set $blob\_uniqueness = 500, topn = 5, name\_similarity = 0.5$, which we find produce satisfactory results.

\subsubsection{Evaluation}
We evaluate the algorithm on the 12,375 correct links.\footnote{For each link, we use the package's latest release before October 2021, the time when the WoC version U data was collected.} The algorithm can successfully retrieve repository information from WoC for 11,165 releases (i.e., coverage: 0.902) with an accuracy of 0.970.
Table~\ref{tab: name similarity} shows the retrieval accuracy and coverage under different $name\_similarity$ setting.
When setting $name\_similarity$ as 1, the accuracy increases to 0.986 but the coverage decreases to 0.757; when setting $name\_similarity$ as 0, the coverage increases to 0.986 but the accuracy decreases to 0.930. Therefore, we set $name\_similarity$ as 0.5 to strike a balance between retrieval coverage and accuracy.
We run this component on the rest releases for which the Metadata-based Retriever can not retrieve repositories. We only select the latest release before October 2021 for each package, resulting in 81,751 releases. This component successfully retrieves repository information for 32,139 (39.3\%) releases from WoC. The relatively low ratio may be attributed to the fact that many releases have repositories that are either not public or not indexed by WoC.
We manually inspect 100 releases and find that this component correctly retrieves repository information for 90 releases, yielding an accuracy of 90\%. On average, it takes 40 seconds to complete a repository retrieval, suggesting its high efficiency.

The Metadata-based Retriever and Source Code-based Retriever retrieve repository information using the release's metadata and source distribution, respectively. Therefore, we compare their retrieval results to further validate the two components. Specifically, the Metadata-based Retriever finds repository URLs for 3,047,112 releases of 273,273 packages. To conduct the comparison, we first select the most recent release before October 2021 (the time of WoC U version data collection) for each package. Then, we keep releases that 1) provide a source distribution for the Source Code-based Retriever to use and 2) have their repository retrieved by the Metadata-based Retriever indexed by WoC. Finally, we obtain 173,809 releases. The Source Code-based Retriever successfully retrieves repository URLs for 143,035 (82.3\%) of the selected releases, with 131,710 (92.1\%) the same as the repository URLs retrieved by the Metadata-based Retriever, indicating a high consistency between the results of the two components.

\begin{table}
\small
    \caption{Retrieval coverage and accuracy under different $name\_similarity$ settings.}
  \label{tab: name similarity}
  \renewcommand{\arraystretch}{1.2}
  \vspace{-3mm}
  \begin{tabular}{lrrrrrrrrrrr}
    \toprule
     & 0.0 &  0.1 & 0.2 & 0.3 & 0.4 & 0.5 & 0.6 & 0.7 & 0.8 & 0.9 & 1.0 \\
     \midrule
     Coverage & 0.986 & 0.985 & 0.976 & 0.954 & 0.927 & 0.902 & 0.876 & 0.843 & 0.803 & 0.768 & 0.757 \\
     Accuracy & 0.930 & 0.932 & 0.939 & 0.954 & 0.965 & 0.970 & 0.975 & 0.979 & 0.982 & 0.985 & 0.986 \\
    \bottomrule
    \end{tabular}
    \vspace{-3mm}
\end{table}

\subsection{Overall Evaluation}

We evaluate PyRadar on the Package-Repository Link dataset with 14,375 correct and 2,064 incorrect links. For a release in the correct link, the expected output is the repository URL in the link. For a release in the incorrect link, the expected output is empty since there is no sufficient information to manually pinpoint its exact repository. As a result, PyRadar achieves an accuracy of 0.88.

\section{Limitations}

Several limitations pertain to the Package-Repository Link dataset. First, we rely on the black-box GitHub Dependency Graph to collect correct links. Therefore, we can not guarantee the accuracy of GDG data.
To alleviate this threat, we carefully select popular Python repositories and conduct a manual evaluation to confirm the validity of the collected data.
Second, the correct links are skewed towards popular packages and popular repositories, which may compromise the representativeness of the data. However, considering the number (14 375, 3.4\% of all PyPI packages) and the monthly downloads (ranging from 28 to 1.02 million) of the collected packages, we believe the collected data are sufficiently representative.
Third, we assume a link is incorrect where the package is linked to a repository that is the same as the repository in the correct link but with a different maintainer.
The assumption has not been systematically tested and may threaten the validity of the collected data. We conduct a manual evaluation of the collected data to alleviate this threat.

In terms of the empirical study, its main threat is the reimplementation of the four tools. To alleviate this threat, we borrow test cases from the original tools to ensure the consistency of the reimplemented tools with the original ones. Also, due to the complexity of Git, the implementation of the repository traversal algorithm may threaten the internal validity. We use as many git native operations as possible to ensure the correctness of the implementation. Therefore, we believe the empirical study is conducted on a solid basis of correct implementation.

Despite the three most popular code hosting platforms considered by the Metadata-based Retriever, the package's source code repository may be hosted on other platforms such as SourceForge and other self-hosted platforms, e.g., \href{https://gitlab.inria.fr/}{GitLab at Inria}. However, this component can be easily extended by adding new regular expressions to match repositories on these platforms.
The major limitation of the Source Code-based Retriever is the dependence on the external infrastructure, WoC. Despite the relatively complete collection of open source Git repositories in WoC, it is impossible for WoC to contain all repositories.
Besides, the several-month update delay of WoC means that the Source Code-based Retriever can only be run periodically to retrieve repository information for releases uploaded to PyPI during the update interval. Despite these limitations, we believe WoC is still the most suitable infrastructure for our tool.
Finally, \toolname~ fails to retrieve a package's repository when 1) the package's repository is not public, e.g., hosted locally or privately. 2) the package's repository is public but is neither declared in the metadata nor indexed by WoC.

Due to the different usage scenarios and the challenge of automatically and accurately collecting ground truth, i.e., correct and incorrect package-repository links, the components are evaluated on different datasets, which may induce the overfitting issue. To evaluate the generalization of the second and third components, we run the second component against 228,448 package-repository links from the first component and run the third component against 81,751 releases for which the first component fails to retrieve repository information, which align with their practical usage scenarios. The manual inspection reveals that the two components achieve an accuracy of 85\% and 90\% respectively, suggesting the overfitting issue induced by different evaluation datasets is minor.

In terms of external validity, the dataset collection approach is dedicated to PyPI. We believe the correct link collection part can be generalized to other package registries since GDG supports multiple packaging tools. However, the incorrect link collection needs further research due to the differences in the package management practices adopted by different package registries~\cite{Ruian2021-NDSS}.
The empirical findings can not be generalized to other package registries, too. However, the repository traversal algorithm can be applied to any repository, laying a foundation for future work on other package registries. Despite the specific design of \toolname~ for PyPI packages, the framework is general and can be adapted to other package registries similar to PyPI (e.g., providing metadata and source distributions).

\section{Discussion}

\subsection{Comparison with Related Work}
The most similar work to ours is~\cite{Sun2023-EMSE}. There are two noteworthy differences between our approach and theirs. First, our approach and~\cite{Sun2023-EMSE} serve different purposes. ~\cite{Sun2023-EMSE} targets the problem of mapping Debian source packages written in Python to PyPI packages, and our approach targets the problem of locating a PyPI package's code repository, which is more intricate due to the sophisticated Git-based software development~\cite{Bird2009-MSR} and a significantly larger retrieval corpus (244 thousand PyPI packages vs. 173 million repositories).

Second, due to different goals, ~\cite{Sun2023-EMSE} and the Source Code-based Retriever in our approach adopt different means but overlap in some particular steps in the whole process. Specifically, ~\cite{Sun2023-EMSE} involves three steps: 1) \textbf{index}: indexing global identifiers (i.e., class names and function names) defined in PyPI packages; 2) \textbf{retrieval}: retrieving candidate PyPI packages with three random identifiers from a random Python file in the Debian package; 3) \textbf{selection}: selecting the most popular candidate using the SourceRank metric. Our Source Code-based Retriever also has the retrieval and selection steps, but is superior in three aspects related to our problem:

\begin{itemize}[leftmargin=*,topsep=0pt]
    \item Our retriever does not require the time-consuming identifier indexing step. The phantom file analysis reveals that Python files are rarely phantom files in correct package-repository links. It suggests that querying candidate repositories with the hashes of Python files in the package is sufficient and does not require the time-consuming identifier indexing step that would \textit{take months for the 12.4 billion blobs in WoC}.
    \item Our retriever utilizes more information to retrieve candidates. First, querying with a Python file's hash, as conducted in our retriever, is equivalent to querying with all identifiers in it, while ~\cite{Sun2023-EMSE} only queries with three identifiers. Second, our retriever uses all Python files, while ~\cite{Sun2023-EMSE} uses only a random Python file.
    \item Our approach selects the final candidate more suitably. First, the SourceRank metric used by~\cite{Sun2023-EMSE} is unsuitable for ranking repositories due to its inclusion of many package-specific factors, such as the presence of a link to the source code. Therefore, the approach in ~\cite{Sun2023-EMSE} does not apply to our problem. Second, ~\cite{Sun2023-EMSE} selects the most popular candidate, whereas our retriever selects the most similar candidate to the package (measured by the number of matched Python files), which better suits our retrieval problem.
\end{itemize}

\subsection{The Feasibility of Name-matching Approach}
Although the package name and repository name are the same in 11,408 (79.36\%) correct links, the name-matching approach is still insufficient for validating and locating a release's repository for three reasons.
First, it is common that a package name differs from its repository name, as evidenced by the 20.64\% of correct links, where the name-matching approach fails to retrieve the correct repository as a candidate.
Second, even if the correct repository is retrieved as a candidate, the presence of many repositories with the same name necessitates further selection of the correct one. Specifically, the median number of repositories in WoC that have the same name as packages in the 14,375 correct links is 24.
Third, when validating the correctness of the remaining 5,031 links, the AUC of using only the \Code{name\_similarity} feature is 0.563, while using all six features yields an AUC of 0.979, suggesting the effectiveness and necessity of the remaining five features.

\subsection{Implication}
Our research highlights the following future improvements in retrieving and validating the release's repository information.

\textit{Facilitate account linking between code hosting platforms and package registries.} Cross-linking accounts can largely alleviate the problem of unavailable or incorrect repository information. On the one hand, package registries and code hosting platforms may collaboratively establish account binding or account authorization login mechanisms. On the other hand, to the best of our knowledge, account cross-linking tools between code hosting platforms and package registries are still lacking. Future work can bridge this gap.

\textit{Alert users to potentially incorrect repository information of the release.} Package registries can integrate repository information validation mechanisms and display validation results on the package's PyPI page and in the metadata, which can be consumed by package managers and related package monitoring tools to alert users when searching for, assessing, or installing packages.

\textit{Conduct code analysis to identify package names built from a repository.} As discussed in Section~\ref{s5.2.2}, our validator does not perform well in cases where the package name differs from the repository name greatly. To alleviate this issue, future work may explore code analysis techniques to precisely parse the package names built from the repository, which will benefit both the validation and retrieval of the release's repository information.

\section{Conclusion}
A package's source code repository is critical for the use and risk monitoring of the package. However, the package's metadata may not contain or contain wrong repository information. In this paper, we collect 4,227,425 PyPI releases' metadata, 14,375 correct package-repository links, and 2,064 incorrect links. Then we systematically compare four existing metadata-based tools and investigate phantom file differences between correct and incorrect links. Inspired by the empirical findings, we propose \toolname, a novel framework that utilizes the PyPI release's metadata and source distribution to automatically retrieve and validate the release's repository information. We believe our work can help both practitioners and researchers better use PyPI packages. We provide a replication package at \url{https://github.com/gaokai320/PyRadar}.

\begin{acks}
This work is sponsored by the National Natural Science Foundation of China 61825201 and 62332001.
\end{acks}

\bibliographystyle{ACM-Reference-Format}
\bibliography{reference}

%%% -*-BibTeX-*-
%%% Do NOT edit. File created by BibTeX with style
%%% ACM-Reference-Format-Journals [18-Jan-2012].

\begin{thebibliography}{73}

%%% ====================================================================
%%% NOTE TO THE USER: you can override these defaults by providing
%%% customized versions of any of these macros before the \bibliography
%%% command.  Each of them MUST provide its own final punctuation,
%%% except for \shownote{}, \showDOI{}, and \showURL{}.  The latter two
%%% do not use final punctuation, in order to avoid confusing it with
%%% the Web address.
%%%
%%% To suppress output of a particular field, define its macro to expand
%%% to an empty string, or better, \unskip, like this:
%%%
%%% \newcommand{\showDOI}[1]{\unskip}   % LaTeX syntax
%%%
%%% \def \showDOI #1{\unskip}           % plain TeX syntax
%%%
%%% ====================================================================

\ifx \showCODEN    \undefined \def \showCODEN     #1{\unskip}     \fi
\ifx \showDOI      \undefined \def \showDOI       #1{#1}\fi
\ifx \showISBNx    \undefined \def \showISBNx     #1{\unskip}     \fi
\ifx \showISBNxiii \undefined \def \showISBNxiii  #1{\unskip}     \fi
\ifx \showISSN     \undefined \def \showISSN      #1{\unskip}     \fi
\ifx \showLCCN     \undefined \def \showLCCN      #1{\unskip}     \fi
\ifx \shownote     \undefined \def \shownote      #1{#1}          \fi
\ifx \showarticletitle \undefined \def \showarticletitle #1{#1}   \fi
\ifx \showURL      \undefined \def \showURL       {\relax}        \fi
% The following commands are used for tagged output and should be
% invisible to TeX
\providecommand\bibfield[2]{#2}
\providecommand\bibinfo[2]{#2}
\providecommand\natexlab[1]{#1}
\providecommand\showeprint[2][]{arXiv:#2}

\bibitem[Alfadel et~al\mbox{.}(2021)]%
        {Alfadel2021-SANER}
\bibfield{author}{\bibinfo{person}{Mahmoud Alfadel}, \bibinfo{person}{Diego~Elias Costa}, {and} \bibinfo{person}{Emad Shihab}.} \bibinfo{year}{2021}\natexlab{}.
\newblock \showarticletitle{Empirical Analysis of Security Vulnerabilities in Python Packages}. In \bibinfo{booktitle}{\emph{2021 IEEE International Conference on Software Analysis, Evolution and Reengineering (SANER)}}. \bibinfo{pages}{446--457}.
\newblock
\urldef\tempurl%
\url{https://doi.org/10.1109/SANER50967.2021.00048}
\showDOI{\tempurl}


\bibitem[Bird et~al\mbox{.}(2009)]%
        {Bird2009-MSR}
\bibfield{author}{\bibinfo{person}{Christian Bird}, \bibinfo{person}{Peter~C. Rigby}, \bibinfo{person}{Earl~T. Barr}, \bibinfo{person}{David~J. Hamilton}, \bibinfo{person}{Daniel~M. German}, {and} \bibinfo{person}{Prem Devanbu}.} \bibinfo{year}{2009}\natexlab{}.
\newblock \showarticletitle{The promises and perils of mining git}. In \bibinfo{booktitle}{\emph{2009 6th IEEE International Working Conference on Mining Software Repositories}}. \bibinfo{pages}{1--10}.
\newblock
\urldef\tempurl%
\url{https://doi.org/10.1109/MSR.2009.5069475}
\showDOI{\tempurl}


\bibitem[Borges et~al\mbox{.}(2016)]%
        {Borges2016-ICSME}
\bibfield{author}{\bibinfo{person}{Hudson Borges}, \bibinfo{person}{Andre Hora}, {and} \bibinfo{person}{Marco~Tulio Valente}.} \bibinfo{year}{2016}\natexlab{}.
\newblock \showarticletitle{Understanding the Factors That Impact the Popularity of GitHub Repositories}. In \bibinfo{booktitle}{\emph{2016 IEEE International Conference on Software Maintenance and Evolution (ICSME)}}. \bibinfo{pages}{334--344}.
\newblock
\urldef\tempurl%
\url{https://doi.org/10.1109/ICSME.2016.31}
\showDOI{\tempurl}


\bibitem[Breiman(2001)]%
        {Breiman2001-ML}
\bibfield{author}{\bibinfo{person}{Leo Breiman}.} \bibinfo{year}{2001}\natexlab{}.
\newblock \showarticletitle{Random Forests}.
\newblock \bibinfo{journal}{\emph{Machine Learning}} \bibinfo{volume}{45}, \bibinfo{number}{1} (\bibinfo{date}{01 Oct} \bibinfo{year}{2001}), \bibinfo{pages}{5--32}.
\newblock
\showISSN{1573-0565}
\urldef\tempurl%
\url{https://doi.org/10.1023/A:1010933404324}
\showDOI{\tempurl}


\bibitem[Chacon and Straub(2023)]%
        {Gitobjects}
\bibfield{author}{\bibinfo{person}{Scott Chacon} {and} \bibinfo{person}{Ben Straub}.} \bibinfo{year}{2023}\natexlab{}.
\newblock \bibinfo{title}{Git - Git Objects}.
\newblock \bibinfo{howpublished}{\url{https://git-scm.com/book/en/v2/Git-Internals-Git-Objects}}.
\newblock
\newblock
\shownote{(Accessed on 09/14/2023)}.


\bibitem[Chen and Guestrin(2016)]%
        {DBLP:conf/kdd/ChenG16}
\bibfield{author}{\bibinfo{person}{Tianqi Chen} {and} \bibinfo{person}{Carlos Guestrin}.} \bibinfo{year}{2016}\natexlab{}.
\newblock \showarticletitle{XGBoost: {A} Scalable Tree Boosting System}. In \bibinfo{booktitle}{\emph{Proceedings of the 22nd {ACM} {SIGKDD} International Conference on Knowledge Discovery and Data Mining, San Francisco, CA, USA, August 13-17, 2016}}, \bibfield{editor}{\bibinfo{person}{Balaji Krishnapuram}, \bibinfo{person}{Mohak Shah}, \bibinfo{person}{Alexander~J. Smola}, \bibinfo{person}{Charu~C. Aggarwal}, \bibinfo{person}{Dou Shen}, {and} \bibinfo{person}{Rajeev Rastogi}} (Eds.). \bibinfo{publisher}{{ACM}}, \bibinfo{pages}{785--794}.
\newblock
\urldef\tempurl%
\url{https://doi.org/10.1145/2939672.2939785}
\showDOI{\tempurl}


\bibitem[Cosmo and Zacchiroli(2017)]%
        {Cosmo2017-iPRES}
\bibfield{author}{\bibinfo{person}{Roberto~Di Cosmo} {and} \bibinfo{person}{Stefano Zacchiroli}.} \bibinfo{year}{2017}\natexlab{}.
\newblock \showarticletitle{Software Heritage: Why and How to Preserve Software Source Code}. In \bibinfo{booktitle}{\emph{Proceedings of the 14th International Conference on Digital Preservation, iPRES 2017, Kyoto, Japan, September 25-29, 2017}}, \bibfield{editor}{\bibinfo{person}{Shoichiro Hara}, \bibinfo{person}{Shigeo Sugimoto}, {and} \bibinfo{person}{Makoto Goto}} (Eds.).
\newblock
\urldef\tempurl%
\url{https://hdl.handle.net/11353/10.931064}
\showURL{%
\tempurl}


\bibitem[coursera\textendash dl(2016)]%
        {coursera-dl}
\bibfield{author}{\bibinfo{person}{coursera\textendash dl}.} \bibinfo{year}{2016}\natexlab{}.
\newblock \bibinfo{title}{Rename PyPI package name from "coursera" to "coursera-dl" · coursera-dl/coursera-dl@c2f318a}.
\newblock \bibinfo{howpublished}{\url{https://github.com/coursera-dl/coursera-dl/commit/c2f318a57183800a8fb9360761651690d7db3e5a}}.
\newblock
\newblock
\shownote{(Accessed on 04/25/2024)}.


\bibitem[Dabbish et~al\mbox{.}(2012)]%
        {Dabbish2012-CSCW}
\bibfield{author}{\bibinfo{person}{Laura Dabbish}, \bibinfo{person}{Colleen Stuart}, \bibinfo{person}{Jason Tsay}, {and} \bibinfo{person}{Jim Herbsleb}.} \bibinfo{year}{2012}\natexlab{}.
\newblock \showarticletitle{Social Coding in GitHub: Transparency and Collaboration in an Open Software Repository}. In \bibinfo{booktitle}{\emph{Proceedings of the ACM 2012 Conference on Computer Supported Cooperative Work}} (Seattle, Washington, USA) \emph{(\bibinfo{series}{CSCW '12})}. \bibinfo{publisher}{Association for Computing Machinery}, \bibinfo{address}{New York, NY, USA}, \bibinfo{pages}{1277–1286}.
\newblock
\showISBNx{9781450310864}
\urldef\tempurl%
\url{https://doi.org/10.1145/2145204.2145396}
\showDOI{\tempurl}


\bibitem[Decan et~al\mbox{.}(2018)]%
        {Decan2018-MSR}
\bibfield{author}{\bibinfo{person}{Alexandre Decan}, \bibinfo{person}{Tom Mens}, {and} \bibinfo{person}{Eleni Constantinou}.} \bibinfo{year}{2018}\natexlab{}.
\newblock \showarticletitle{On the Impact of Security Vulnerabilities in the Npm Package Dependency Network}. In \bibinfo{booktitle}{\emph{Proceedings of the 15th International Conference on Mining Software Repositories}} (Gothenburg, Sweden) \emph{(\bibinfo{series}{MSR '18})}. \bibinfo{publisher}{Association for Computing Machinery}, \bibinfo{address}{New York, NY, USA}, \bibinfo{pages}{181–191}.
\newblock
\showISBNx{9781450357166}
\urldef\tempurl%
\url{https://doi.org/10.1145/3196398.3196401}
\showDOI{\tempurl}


\bibitem[Duan et~al\mbox{.}(2021)]%
        {Ruian2021-NDSS}
\bibfield{author}{\bibinfo{person}{Ruian Duan}, \bibinfo{person}{Omar Alrawi}, \bibinfo{person}{Ranjita~Pai Kasturi}, \bibinfo{person}{Ryan Elder}, \bibinfo{person}{Brendan Saltaformaggio}, {and} \bibinfo{person}{Wenke Lee}.} \bibinfo{year}{2021}\natexlab{}.
\newblock \showarticletitle{Towards Measuring Supply Chain Attacks on Package Managers for Interpreted Languages}. In \bibinfo{booktitle}{\emph{28th Annual Network and Distributed System Security Symposium, {NDSS} 2021, virtually, February 21-25, 2021}}. \bibinfo{publisher}{The Internet Society}.
\newblock
\urldef\tempurl%
\url{https://www.ndss-symposium.org/ndss-paper/towards-measuring-supply-chain-attacks-on-package-managers-for-interpreted-languages/}
\showURL{%
\tempurl}


\bibitem[Edward2(2019)]%
        {Edward2}
\bibfield{author}{\bibinfo{person}{Edward2}.} \bibinfo{year}{2019}\natexlab{}.
\newblock \bibinfo{title}{Move Edward2 from google-research/google-research to google/edward2. · google-research/google-research@f26db54}.
\newblock \bibinfo{howpublished}{\url{https://github.com/google-research/google-research/commit/f26db5490fa147a6052a78b2e479361833c3fd41}}.
\newblock
\newblock
\shownote{(Accessed on 04/25/2024)}.


\bibitem[Fang et~al\mbox{.}(2020)]%
        {Fang2020-MSR}
\bibfield{author}{\bibinfo{person}{Hongbo Fang}, \bibinfo{person}{Daniel Klug}, \bibinfo{person}{Hemank Lamba}, \bibinfo{person}{James Herbsleb}, {and} \bibinfo{person}{Bogdan Vasilescu}.} \bibinfo{year}{2020}\natexlab{}.
\newblock \showarticletitle{Need for Tweet: How Open Source Developers Talk About Their GitHub Work on Twitter}. In \bibinfo{booktitle}{\emph{Proceedings of the 17th International Conference on Mining Software Repositories}} (Seoul, Republic of Korea) \emph{(\bibinfo{series}{MSR '20})}. \bibinfo{publisher}{Association for Computing Machinery}, \bibinfo{address}{New York, NY, USA}, \bibinfo{pages}{322–326}.
\newblock
\showISBNx{9781450375177}
\urldef\tempurl%
\url{https://doi.org/10.1145/3379597.3387466}
\showDOI{\tempurl}


\bibitem[Foundation(2023)]%
        {PEP527}
\bibfield{author}{\bibinfo{person}{Python~Software Foundation}.} \bibinfo{year}{2023}\natexlab{}.
\newblock \bibinfo{title}{PEP 527 – Removing Un(der)used file types/extensions on PyPI | peps.python.org}.
\newblock \bibinfo{howpublished}{\url{https://peps.python.org/pep-0527/}}.
\newblock
\newblock
\shownote{(Accessed on 09/24/2023)}.


\bibitem[Freund and Schapire(1997)]%
        {freund1997decision}
\bibfield{author}{\bibinfo{person}{Yoav Freund} {and} \bibinfo{person}{Robert~E Schapire}.} \bibinfo{year}{1997}\natexlab{}.
\newblock \showarticletitle{A decision-theoretic generalization of on-line learning and an application to boosting}.
\newblock \bibinfo{journal}{\emph{Journal of computer and system sciences}} \bibinfo{volume}{55}, \bibinfo{number}{1} (\bibinfo{year}{1997}), \bibinfo{pages}{119--139}.
\newblock


\bibitem[Friedman(2001)]%
        {friedman2001greedy}
\bibfield{author}{\bibinfo{person}{Jerome~H Friedman}.} \bibinfo{year}{2001}\natexlab{}.
\newblock \showarticletitle{Greedy function approximation: a gradient boosting machine}.
\newblock \bibinfo{journal}{\emph{Annals of statistics}} (\bibinfo{year}{2001}), \bibinfo{pages}{1189--1232}.
\newblock


\bibitem[GitHub(2023a)]%
        {GitHubDependencyGraph}
\bibfield{author}{\bibinfo{person}{GitHub}.} \bibinfo{year}{2023}\natexlab{a}.
\newblock \bibinfo{title}{About the dependency graph - GitHub Docs}.
\newblock \bibinfo{howpublished}{\url{https://docs.github.com/en/code-security/supply-chain-security/understanding-your-software-supply-chain/about-the-dependency-graph}}.
\newblock
\newblock
\shownote{(Accessed on 09/21/2023)}.


\bibitem[GitHub(2023b)]%
        {GitHubSearchAPI}
\bibfield{author}{\bibinfo{person}{GitHub}.} \bibinfo{year}{2023}\natexlab{b}.
\newblock \bibinfo{title}{Search - GitHub Docs}.
\newblock \bibinfo{howpublished}{\url{https://docs.github.com/en/free-pro-team@latest/rest/search/search?apiVersion=2022-11-28\#search-repositories}}.
\newblock
\newblock
\shownote{(Accessed on 09/21/2023)}.


\bibitem[Godfrey(2015)]%
        {Godfrey2015-SCP}
\bibfield{author}{\bibinfo{person}{Michael~W. Godfrey}.} \bibinfo{year}{2015}\natexlab{}.
\newblock \showarticletitle{Understanding software artifact provenance}.
\newblock \bibinfo{journal}{\emph{Science of Computer Programming}}  \bibinfo{volume}{97} (\bibinfo{year}{2015}), \bibinfo{pages}{86--90}.
\newblock
\showISSN{0167-6423}
\urldef\tempurl%
\url{https://doi.org/10.1016/j.scico.2013.11.021}
\showDOI{\tempurl}
\newblock
\shownote{Special Issue on New Ideas and Emerging Results in Understanding Software}.


\bibitem[Google(2021a)]%
        {BigQuery88:online}
\bibfield{author}{\bibinfo{person}{Google}.} \bibinfo{year}{2021}\natexlab{a}.
\newblock \bibinfo{title}{BigQuery dataset | Open Source Insights}.
\newblock \bibinfo{howpublished}{\url{https://docs.deps.dev/bigquery/v1/\#packageversiontoproject}}.
\newblock
\newblock
\shownote{(Accessed on 04/24/2024)}.


\bibitem[Google(2021b)]%
        {Frequent71:online}
\bibfield{author}{\bibinfo{person}{Google}.} \bibinfo{year}{2021}\natexlab{b}.
\newblock \bibinfo{title}{Frequently Asked Questions | Open Source Insights}.
\newblock \bibinfo{howpublished}{\url{https://docs.deps.dev/faq/}}.
\newblock
\newblock
\shownote{(Accessed on 04/24/2024)}.


\bibitem[Google(2021c)]%
        {OpenSourceInsights}
\bibfield{author}{\bibinfo{person}{Google}.} \bibinfo{year}{2021}\natexlab{c}.
\newblock \bibinfo{title}{Open Source Insights}.
\newblock \bibinfo{howpublished}{\url{https://deps.dev/}}.
\newblock
\newblock
\shownote{(Accessed on 09/01/2023)}.


\bibitem[Hanley et~al\mbox{.}(1982)]%
        {hanley1982use}
\bibfield{author}{\bibinfo{person}{James~A Hanley}, \bibinfo{person}{Barbara~J Mc~Neil}, {and} \bibinfo{person}{A Meaning}.} \bibinfo{year}{1982}\natexlab{}.
\newblock \showarticletitle{use of the area under a receiver Operating Characteristics (ROC) curves}.
\newblock \bibinfo{journal}{\emph{Radiology}} \bibinfo{volume}{143}, \bibinfo{number}{1} (\bibinfo{year}{1982}), \bibinfo{pages}{29--36}.
\newblock


\bibitem[Hata et~al\mbox{.}(2021)]%
        {Hata2021-ICSE}
\bibfield{author}{\bibinfo{person}{Hideaki Hata}, \bibinfo{person}{Raula~Gaikovina Kula}, \bibinfo{person}{Takashi Ishio}, {and} \bibinfo{person}{Christoph Treude}.} \bibinfo{year}{2021}\natexlab{}.
\newblock \showarticletitle{Same File, Different Changes: The Potential of Meta-Maintenance on GitHub}. In \bibinfo{booktitle}{\emph{2021 IEEE/ACM 43rd International Conference on Software Engineering (ICSE)}}. \bibinfo{pages}{773--784}.
\newblock
\urldef\tempurl%
\url{https://doi.org/10.1109/ICSE43902.2021.00076}
\showDOI{\tempurl}


\bibitem[He et~al\mbox{.}(2021)]%
        {He2021-FSE}
\bibfield{author}{\bibinfo{person}{Hao He}, \bibinfo{person}{Runzhi He}, \bibinfo{person}{Haiqiao Gu}, {and} \bibinfo{person}{Minghui Zhou}.} \bibinfo{year}{2021}\natexlab{}.
\newblock \showarticletitle{A Large-Scale Empirical Study on Java Library Migrations: Prevalence, Trends, and Rationales}. In \bibinfo{booktitle}{\emph{Proceedings of the 29th ACM Joint Meeting on European Software Engineering Conference and Symposium on the Foundations of Software Engineering}} (Athens, Greece) \emph{(\bibinfo{series}{ESEC/FSE 2021})}. \bibinfo{publisher}{Association for Computing Machinery}, \bibinfo{address}{New York, NY, USA}, \bibinfo{pages}{478–490}.
\newblock
\showISBNx{9781450385626}
\urldef\tempurl%
\url{https://doi.org/10.1145/3468264.3468571}
\showDOI{\tempurl}


\bibitem[Ladisa et~al\mbox{.}(2023)]%
        {Ladisa2023-SP}
\bibfield{author}{\bibinfo{person}{P. Ladisa}, \bibinfo{person}{H. Plate}, \bibinfo{person}{M. Martinez}, {and} \bibinfo{person}{O. Barais}.} \bibinfo{year}{2023}\natexlab{}.
\newblock \showarticletitle{SoK: Taxonomy of Attacks on Open-Source Software Supply Chains}. In \bibinfo{booktitle}{\emph{2023 IEEE Symposium on Security and Privacy (SP)}}. \bibinfo{publisher}{IEEE Computer Society}, \bibinfo{address}{Los Alamitos, CA, USA}, \bibinfo{pages}{1509--1526}.
\newblock
\urldef\tempurl%
\url{https://doi.org/10.1109/SP46215.2023.10179304}
\showDOI{\tempurl}


\bibitem[Larios~Vargas et~al\mbox{.}(2020)]%
        {Larios2020-FSE}
\bibfield{author}{\bibinfo{person}{Enrique Larios~Vargas}, \bibinfo{person}{Maur\'{\i}cio Aniche}, \bibinfo{person}{Christoph Treude}, \bibinfo{person}{Magiel Bruntink}, {and} \bibinfo{person}{Georgios Gousios}.} \bibinfo{year}{2020}\natexlab{}.
\newblock \showarticletitle{Selecting Third-Party Libraries: The Practitioners’ Perspective}. In \bibinfo{booktitle}{\emph{Proceedings of the 28th ACM Joint Meeting on European Software Engineering Conference and Symposium on the Foundations of Software Engineering}} (Virtual Event, USA) \emph{(\bibinfo{series}{ESEC/FSE 2020})}. \bibinfo{publisher}{Association for Computing Machinery}, \bibinfo{address}{New York, NY, USA}, \bibinfo{pages}{245–256}.
\newblock
\showISBNx{9781450370431}
\urldef\tempurl%
\url{https://doi.org/10.1145/3368089.3409711}
\showDOI{\tempurl}


\bibitem[Levenshtein et~al\mbox{.}(1966)]%
        {levenshtein1966binary}
\bibfield{author}{\bibinfo{person}{Vladimir~I Levenshtein} {et~al\mbox{.}}} \bibinfo{year}{1966}\natexlab{}.
\newblock \showarticletitle{Binary codes capable of correcting deletions, insertions, and reversals}. In \bibinfo{booktitle}{\emph{Soviet physics doklady}}, Vol.~\bibinfo{volume}{10}. Soviet Union, \bibinfo{pages}{707--710}.
\newblock


\bibitem[Liu et~al\mbox{.}(2022)]%
        {Liu2022-ICSE}
\bibfield{author}{\bibinfo{person}{Chengwei Liu}, \bibinfo{person}{Sen Chen}, \bibinfo{person}{Lingling Fan}, \bibinfo{person}{Bihuan Chen}, \bibinfo{person}{Yang Liu}, {and} \bibinfo{person}{Xin Peng}.} \bibinfo{year}{2022}\natexlab{}.
\newblock \showarticletitle{Demystifying the Vulnerability Propagation and Its Evolution via Dependency Trees in the NPM Ecosystem}. In \bibinfo{booktitle}{\emph{Proceedings of the 44th International Conference on Software Engineering}} (Pittsburgh, Pennsylvania) \emph{(\bibinfo{series}{ICSE '22})}. \bibinfo{publisher}{Association for Computing Machinery}, \bibinfo{address}{New York, NY, USA}, \bibinfo{pages}{672–684}.
\newblock
\showISBNx{9781450392211}
\urldef\tempurl%
\url{https://doi.org/10.1145/3510003.3510142}
\showDOI{\tempurl}


\bibitem[Ma et~al\mbox{.}(2019)]%
        {Ma2019-MSR}
\bibfield{author}{\bibinfo{person}{Yuxing Ma}, \bibinfo{person}{Chris Bogart}, \bibinfo{person}{Sadika Amreen}, \bibinfo{person}{Russell Zaretzki}, {and} \bibinfo{person}{Audris Mockus}.} \bibinfo{year}{2019}\natexlab{}.
\newblock \showarticletitle{World of Code: An Infrastructure for Mining the Universe of Open Source VCS Data}. In \bibinfo{booktitle}{\emph{Proceedings of the 16th International Conference on Mining Software Repositories}} (Montreal, Quebec, Canada) \emph{(\bibinfo{series}{MSR '19})}. \bibinfo{publisher}{IEEE Press}, \bibinfo{pages}{143–154}.
\newblock
\urldef\tempurl%
\url{https://doi.org/10.1109/MSR.2019.00031}
\showDOI{\tempurl}


\bibitem[Ma et~al\mbox{.}(2021)]%
        {Ma2021-EMSE}
\bibfield{author}{\bibinfo{person}{Yuxing Ma}, \bibinfo{person}{Tapajit Dey}, \bibinfo{person}{Chris Bogart}, \bibinfo{person}{Sadika Amreen}, \bibinfo{person}{Marat Valiev}, \bibinfo{person}{Adam Tutko}, \bibinfo{person}{David Kennard}, \bibinfo{person}{Russell Zaretzki}, {and} \bibinfo{person}{Audris Mockus}.} \bibinfo{year}{2021}\natexlab{}.
\newblock \showarticletitle{World of Code: Enabling a Research Workflow for Mining and Analyzing the Universe of Open Source VCS Data}.
\newblock \bibinfo{journal}{\emph{Empirical Softw. Engg.}} \bibinfo{volume}{26}, \bibinfo{number}{2} (\bibinfo{date}{mar} \bibinfo{year}{2021}), \bibinfo{numpages}{42}~pages.
\newblock
\showISSN{1382-3256}
\urldef\tempurl%
\url{https://doi.org/10.1007/s10664-020-09905-9}
\showDOI{\tempurl}


\bibitem[Mann and Whitney(1947)]%
        {mannwhitneyu}
\bibfield{author}{\bibinfo{person}{H.~B. Mann} {and} \bibinfo{person}{D.~R. Whitney}.} \bibinfo{year}{1947}\natexlab{}.
\newblock \showarticletitle{{On a Test of Whether one of Two Random Variables is Stochastically Larger than the Other}}.
\newblock \bibinfo{journal}{\emph{The Annals of Mathematical Statistics}} \bibinfo{volume}{18}, \bibinfo{number}{1} (\bibinfo{year}{1947}), \bibinfo{pages}{50 -- 60}.
\newblock
\urldef\tempurl%
\url{https://doi.org/10.1214/aoms/1177730491}
\showDOI{\tempurl}


\bibitem[Melo(2013)]%
        {Melo2013}
\bibfield{author}{\bibinfo{person}{Francisco Melo}.} \bibinfo{year}{2013}\natexlab{}.
\newblock \bibinfo{booktitle}{\emph{Area under the ROC Curve}}.
\newblock \bibinfo{publisher}{Springer New York}, \bibinfo{address}{New York, NY}, \bibinfo{pages}{38--39}.
\newblock
\showISBNx{978-1-4419-9863-7}
\urldef\tempurl%
\url{https://doi.org/10.1007/978-1-4419-9863-7_209}
\showDOI{\tempurl}


\bibitem[Microsoft(2020)]%
        {ossgadget}
\bibfield{author}{\bibinfo{person}{Microsoft}.} \bibinfo{year}{2020}\natexlab{}.
\newblock \bibinfo{title}{microsoft/OSSGadget: Collection of tools for analyzing open source packages.}
\newblock \bibinfo{howpublished}{\url{https://github.com/microsoft/OSSGadget}}.
\newblock
\newblock
\shownote{(Accessed on 09/01/2023)}.


\bibitem[Microsoft(2023)]%
        {OSSFindSource}
\bibfield{author}{\bibinfo{person}{Microsoft}.} \bibinfo{year}{2023}\natexlab{}.
\newblock \bibinfo{title}{OSS Find Source · microsoft/OSSGadget Wiki}.
\newblock \bibinfo{howpublished}{\url{https://github.com/microsoft/OSSGadget/wiki/OSS-Find-Source}}.
\newblock
\newblock
\shownote{(Accessed on 09/20/2023)}.


\bibitem[Nguyen et~al\mbox{.}(2020)]%
        {Phuong2020-JSS}
\bibfield{author}{\bibinfo{person}{Phuong~T. Nguyen}, \bibinfo{person}{Juri {Di Rocco}}, \bibinfo{person}{Davide {Di Ruscio}}, {and} \bibinfo{person}{Massimiliano {Di Penta}}.} \bibinfo{year}{2020}\natexlab{}.
\newblock \showarticletitle{CrossRec: Supporting software developers by recommending third-party libraries}.
\newblock \bibinfo{journal}{\emph{Journal of Systems and Software}}  \bibinfo{volume}{161} (\bibinfo{year}{2020}), \bibinfo{pages}{110460}.
\newblock
\showISSN{0164-1212}
\urldef\tempurl%
\url{https://doi.org/10.1016/j.jss.2019.110460}
\showDOI{\tempurl}


\bibitem[Ohm et~al\mbox{.}(2020)]%
        {Ohm2020-DIMVA}
\bibfield{author}{\bibinfo{person}{Marc Ohm}, \bibinfo{person}{Henrik Plate}, \bibinfo{person}{Arnold Sykosch}, {and} \bibinfo{person}{Michael Meier}.} \bibinfo{year}{2020}\natexlab{}.
\newblock \showarticletitle{Backstabber's Knife Collection: A Review of Open Source Software Supply Chain Attacks}. In \bibinfo{booktitle}{\emph{Detection of Intrusions and Malware, and Vulnerability Assessment}}, \bibfield{editor}{\bibinfo{person}{Cl{\'e}mentine Maurice}, \bibinfo{person}{Leyla Bilge}, \bibinfo{person}{Gianluca Stringhini}, {and} \bibinfo{person}{Nuno Neves}} (Eds.). \bibinfo{publisher}{Springer International Publishing}, \bibinfo{address}{Cham}, \bibinfo{pages}{23--43}.
\newblock
\showISBNx{978-3-030-52683-2}


\bibitem[OpenSSF(2023)]%
        {OpenSSFScorecard}
\bibfield{author}{\bibinfo{person}{OpenSSF}.} \bibinfo{year}{2023}\natexlab{}.
\newblock \bibinfo{title}{OpenSSF Scorecard}.
\newblock \bibinfo{howpublished}{\url{https://securityscorecards.dev/}}.
\newblock
\newblock
\shownote{(Accessed on 09/20/2023)}.


\bibitem[Pan et~al\mbox{.}(2022)]%
        {Pan2022-FSE}
\bibfield{author}{\bibinfo{person}{Shengyi Pan}, \bibinfo{person}{Jiayuan Zhou}, \bibinfo{person}{Filipe~Roseiro Cogo}, \bibinfo{person}{Xin Xia}, \bibinfo{person}{Lingfeng Bao}, \bibinfo{person}{Xing Hu}, \bibinfo{person}{Shanping Li}, {and} \bibinfo{person}{Ahmed~E. Hassan}.} \bibinfo{year}{2022}\natexlab{}.
\newblock \showarticletitle{Automated Unearthing of Dangerous Issue Reports}. In \bibinfo{booktitle}{\emph{Proceedings of the 30th ACM Joint European Software Engineering Conference and Symposium on the Foundations of Software Engineering}} (Singapore, Singapore) \emph{(\bibinfo{series}{ESEC/FSE 2022})}. \bibinfo{publisher}{Association for Computing Machinery}, \bibinfo{address}{New York, NY, USA}, \bibinfo{pages}{834–846}.
\newblock
\showISBNx{9781450394130}
\urldef\tempurl%
\url{https://doi.org/10.1145/3540250.3549156}
\showDOI{\tempurl}


\bibitem[Panichella et~al\mbox{.}(2021)]%
        {Sebastiano2021-IST}
\bibfield{author}{\bibinfo{person}{Sebastiano Panichella}, \bibinfo{person}{Gerardo Canfora}, {and} \bibinfo{person}{Andrea {Di Sorbo}}.} \bibinfo{year}{2021}\natexlab{}.
\newblock \showarticletitle{“Won’t We Fix this Issue?” Qualitative characterization and automated identification of wontfix issues on GitHub}.
\newblock \bibinfo{journal}{\emph{Information and Software Technology}}  \bibinfo{volume}{139} (\bibinfo{year}{2021}), \bibinfo{pages}{106665}.
\newblock
\showISSN{0950-5849}
\urldef\tempurl%
\url{https://doi.org/10.1016/j.infsof.2021.106665}
\showDOI{\tempurl}


\bibitem[Pashchenko et~al\mbox{.}(2018)]%
        {Pashchenko2018-ESEM}
\bibfield{author}{\bibinfo{person}{Ivan Pashchenko}, \bibinfo{person}{Henrik Plate}, \bibinfo{person}{Serena~Elisa Ponta}, \bibinfo{person}{Antonino Sabetta}, {and} \bibinfo{person}{Fabio Massacci}.} \bibinfo{year}{2018}\natexlab{}.
\newblock \showarticletitle{Vulnerable Open Source Dependencies: Counting Those That Matter}. In \bibinfo{booktitle}{\emph{Proceedings of the 12th ACM/IEEE International Symposium on Empirical Software Engineering and Measurement}} (Oulu, Finland) \emph{(\bibinfo{series}{ESEM '18})}. \bibinfo{publisher}{Association for Computing Machinery}, \bibinfo{address}{New York, NY, USA}, Article \bibinfo{articleno}{42}, \bibinfo{numpages}{10}~pages.
\newblock
\showISBNx{9781450358231}
\urldef\tempurl%
\url{https://doi.org/10.1145/3239235.3268920}
\showDOI{\tempurl}


\bibitem[project contributors(2023)]%
        {ansible}
\bibfield{author}{\bibinfo{person}{Ansible project contributors}.} \bibinfo{year}{2023}\natexlab{}.
\newblock \bibinfo{title}{Releases and maintenance — Ansible Documentation}.
\newblock \bibinfo{howpublished}{\url{https://docs.ansible.com/ansible/devel/reference_appendices/release_and_maintenance.html}}.
\newblock
\newblock
\shownote{(Accessed on 09/22/2023)}.


\bibitem[PSF(2023)]%
        {PyPIMaintainer}
\bibfield{author}{\bibinfo{person}{PSF}.} \bibinfo{year}{2023}\natexlab{}.
\newblock \bibinfo{title}{Help · PyPI}.
\newblock \bibinfo{howpublished}{\url{https://pypi.org/help/\#collaborator-roles}}.
\newblock
\newblock
\shownote{(Accessed on 09/21/2023)}.


\bibitem[PyPA(2023a)]%
        {Coremetadata}
\bibfield{author}{\bibinfo{person}{PyPA}.} \bibinfo{year}{2023}\natexlab{a}.
\newblock \bibinfo{title}{Core metadata specifications — Python Packaging User Guide}.
\newblock \bibinfo{howpublished}{\url{https://packaging.python.org/en/latest/specifications/core-metadata/}}.
\newblock
\newblock
\shownote{(Accessed on 09/13/2023)}.


\bibitem[PyPA(2023b)]%
        {Glossary}
\bibfield{author}{\bibinfo{person}{PyPA}.} \bibinfo{year}{2023}\natexlab{b}.
\newblock \bibinfo{title}{Glossary — Python Packaging User Guide}.
\newblock \bibinfo{howpublished}{\url{https://packaging.python.org/en/latest/glossary/}}.
\newblock
\newblock
\shownote{(Accessed on 09/16/2023)}.


\bibitem[PyPA(2023c)]%
        {PackagingUserGuide}
\bibfield{author}{\bibinfo{person}{PyPA}.} \bibinfo{year}{2023}\natexlab{c}.
\newblock \bibinfo{title}{Packaging and distributing projects — Python Packaging User Guide}.
\newblock \bibinfo{howpublished}{\url{https://packaging.python.org/en/latest/guides/distributing-packages-using-setuptools/\#packaging-and-distributing-projects}}.
\newblock
\newblock
\shownote{(Accessed on 09/13/2023)}.


\bibitem[PyPI(2023)]%
        {Warehouse}
\bibfield{author}{\bibinfo{person}{PyPI}.} \bibinfo{year}{2023}\natexlab{}.
\newblock \bibinfo{title}{Warehouse documentation}.
\newblock \bibinfo{howpublished}{\url{https://warehouse.pypa.io/}}.
\newblock
\newblock
\shownote{(Accessed on 09/01/2023)}.


\bibitem[Reid et~al\mbox{.}(2022)]%
        {Reid2022-ICSE}
\bibfield{author}{\bibinfo{person}{David Reid}, \bibinfo{person}{Mahmoud Jahanshahi}, {and} \bibinfo{person}{Audris Mockus}.} \bibinfo{year}{2022}\natexlab{}.
\newblock \showarticletitle{The Extent of Orphan Vulnerabilities from Code Reuse in Open Source Software}. In \bibinfo{booktitle}{\emph{Proceedings of the 44th International Conference on Software Engineering}} (Pittsburgh, Pennsylvania) \emph{(\bibinfo{series}{ICSE '22})}. \bibinfo{publisher}{Association for Computing Machinery}, \bibinfo{address}{New York, NY, USA}, \bibinfo{pages}{2104–2115}.
\newblock
\showISBNx{9781450392211}
\urldef\tempurl%
\url{https://doi.org/10.1145/3510003.3510216}
\showDOI{\tempurl}


\bibitem[Rousseau et~al\mbox{.}(2020)]%
        {Rousseau2020-EMSE}
\bibfield{author}{\bibinfo{person}{Guillaume Rousseau}, \bibinfo{person}{Roberto Di~Cosmo}, {and} \bibinfo{person}{Stefano Zacchiroli}.} \bibinfo{year}{2020}\natexlab{}.
\newblock \showarticletitle{Software provenance tracking at the scale of public source code}.
\newblock \bibinfo{journal}{\emph{Empirical Software Engineering}} \bibinfo{volume}{25}, \bibinfo{number}{4} (\bibinfo{date}{01 Jul} \bibinfo{year}{2020}), \bibinfo{pages}{2930--2959}.
\newblock
\showISSN{1573-7616}
\urldef\tempurl%
\url{https://doi.org/10.1007/s10664-020-09828-5}
\showDOI{\tempurl}


\bibitem[Scott and Ben(2023)]%
        {GitSubmodules}
\bibfield{author}{\bibinfo{person}{Chacon Scott} {and} \bibinfo{person}{Straub Ben}.} \bibinfo{year}{2023}\natexlab{}.
\newblock \bibinfo{title}{Git - Submodules}.
\newblock \bibinfo{howpublished}{\url{https://git-scm.com/book/en/v2/Git-Tools-Submodules}}.
\newblock
\newblock
\shownote{(Accessed on 09/19/2023)}.


\bibitem[Silvestri et~al\mbox{.}(2015)]%
        {Silvestri2015-KDWeb}
\bibfield{author}{\bibinfo{person}{Giuseppe Silvestri}, \bibinfo{person}{Jie Yang}, \bibinfo{person}{Alessandro Bozzon}, \bibinfo{person}{Andrea Tagarelli}, {et~al\mbox{.}}} \bibinfo{year}{2015}\natexlab{}.
\newblock \showarticletitle{Linking Accounts across Social Networks: the Case of StackOverflow, Github and Twitter.}. In \bibinfo{booktitle}{\emph{KDWeb}}. \bibinfo{pages}{41--52}.
\newblock


\bibitem[Snyk(2023)]%
        {SnykAdvisor}
\bibfield{author}{\bibinfo{person}{Snyk}.} \bibinfo{year}{2023}\natexlab{}.
\newblock \bibinfo{title}{Snyk Open Source Advisor | Snyk}.
\newblock \bibinfo{howpublished}{\url{https://snyk.io/advisor/python}}.
\newblock
\newblock
\shownote{(Accessed on 09/01/2023)}.


\bibitem[Sun et~al\mbox{.}(2023)]%
        {Sun2023-EMSE}
\bibfield{author}{\bibinfo{person}{Yiming Sun}, \bibinfo{person}{Daniel German}, {and} \bibinfo{person}{Stefano Zacchiroli}.} \bibinfo{year}{2023}\natexlab{}.
\newblock \showarticletitle{Using the uniqueness of global identifiers to determine the provenance of Python software source code}.
\newblock \bibinfo{journal}{\emph{Empirical Software Engineering}} \bibinfo{volume}{28}, \bibinfo{number}{5} (\bibinfo{date}{20 Jul} \bibinfo{year}{2023}), \bibinfo{pages}{107}.
\newblock
\showISSN{1573-7616}
\urldef\tempurl%
\url{https://doi.org/10.1007/s10664-023-10317-8}
\showDOI{\tempurl}


\bibitem[swsc(2023)]%
        {wocoverview}
\bibfield{author}{\bibinfo{person}{swsc}.} \bibinfo{year}{2023}\natexlab{}.
\newblock \bibinfo{title}{swsc / overview — Bitbucket}.
\newblock \bibinfo{howpublished}{\url{https://bitbucket.org/swsc/overview/src/master/}}.
\newblock
\newblock
\shownote{(Accessed on 09/18/2023)}.


\bibitem[Taylor et~al\mbox{.}(2020)]%
        {Taylor2020-NSS}
\bibfield{author}{\bibinfo{person}{Matthew Taylor}, \bibinfo{person}{Ruturaj Vaidya}, \bibinfo{person}{Drew Davidson}, \bibinfo{person}{Lorenzo De~Carli}, {and} \bibinfo{person}{Vaibhav Rastogi}.} \bibinfo{year}{2020}\natexlab{}.
\newblock \showarticletitle{Defending Against Package Typosquatting}. In \bibinfo{booktitle}{\emph{Network and System Security}}, \bibfield{editor}{\bibinfo{person}{Miros{\l}aw Kuty{\l}owski}, \bibinfo{person}{Jun Zhang}, {and} \bibinfo{person}{Chao Chen}} (Eds.). \bibinfo{publisher}{Springer International Publishing}, \bibinfo{address}{Cham}, \bibinfo{pages}{112--131}.
\newblock
\showISBNx{978-3-030-65745-1}


\bibitem[Tian et~al\mbox{.}(2022)]%
        {Tian2022-ICSE}
\bibfield{author}{\bibinfo{person}{Yingchen Tian}, \bibinfo{person}{Yuxia Zhang}, \bibinfo{person}{Klaas-Jan Stol}, \bibinfo{person}{Lin Jiang}, {and} \bibinfo{person}{Hui Liu}.} \bibinfo{year}{2022}\natexlab{}.
\newblock \showarticletitle{What Makes a Good Commit Message?}. In \bibinfo{booktitle}{\emph{Proceedings of the 44th International Conference on Software Engineering}} (Pittsburgh, Pennsylvania) \emph{(\bibinfo{series}{ICSE '22})}. \bibinfo{publisher}{Association for Computing Machinery}, \bibinfo{address}{New York, NY, USA}, \bibinfo{pages}{2389–2401}.
\newblock
\showISBNx{9781450392211}
\urldef\tempurl%
\url{https://doi.org/10.1145/3510003.3510205}
\showDOI{\tempurl}


\bibitem[Tidelift(2015)]%
        {Librariesio}
\bibfield{author}{\bibinfo{person}{Tidelift}.} \bibinfo{year}{2015}\natexlab{}.
\newblock \bibinfo{title}{Libraries.io - The Open Source Discovery Service}.
\newblock \bibinfo{howpublished}{\url{https://libraries.io/}}.
\newblock
\newblock
\shownote{(Accessed on 09/01/2023)}.


\bibitem[Tsay et~al\mbox{.}(2014)]%
        {Tsay2014-ICSE}
\bibfield{author}{\bibinfo{person}{Jason Tsay}, \bibinfo{person}{Laura Dabbish}, {and} \bibinfo{person}{James Herbsleb}.} \bibinfo{year}{2014}\natexlab{}.
\newblock \showarticletitle{Influence of Social and Technical Factors for Evaluating Contribution in GitHub}. In \bibinfo{booktitle}{\emph{Proceedings of the 36th International Conference on Software Engineering}} (Hyderabad, India) \emph{(\bibinfo{series}{ICSE 2014})}. \bibinfo{publisher}{Association for Computing Machinery}, \bibinfo{address}{New York, NY, USA}, \bibinfo{pages}{356–366}.
\newblock
\showISBNx{9781450327565}
\urldef\tempurl%
\url{https://doi.org/10.1145/2568225.2568315}
\showDOI{\tempurl}


\bibitem[Valiev et~al\mbox{.}(2018)]%
        {Valiev2018-FSE}
\bibfield{author}{\bibinfo{person}{Marat Valiev}, \bibinfo{person}{Bogdan Vasilescu}, {and} \bibinfo{person}{James Herbsleb}.} \bibinfo{year}{2018}\natexlab{}.
\newblock \showarticletitle{Ecosystem-Level Determinants of Sustained Activity in Open-Source Projects: A Case Study of the PyPI Ecosystem}. In \bibinfo{booktitle}{\emph{Proceedings of the 2018 26th ACM Joint Meeting on European Software Engineering Conference and Symposium on the Foundations of Software Engineering}} (Lake Buena Vista, FL, USA) \emph{(\bibinfo{series}{ESEC/FSE 2018})}. \bibinfo{publisher}{Association for Computing Machinery}, \bibinfo{address}{New York, NY, USA}, \bibinfo{pages}{644–655}.
\newblock
\showISBNx{9781450355735}
\urldef\tempurl%
\url{https://doi.org/10.1145/3236024.3236062}
\showDOI{\tempurl}


\bibitem[Vasilescu et~al\mbox{.}(2013)]%
        {Vasilescu2013-ICSC}
\bibfield{author}{\bibinfo{person}{Bogdan Vasilescu}, \bibinfo{person}{Vladimir Filkov}, {and} \bibinfo{person}{Alexander Serebrenik}.} \bibinfo{year}{2013}\natexlab{}.
\newblock \showarticletitle{StackOverflow and GitHub: Associations between Software Development and Crowdsourced Knowledge}. In \bibinfo{booktitle}{\emph{2013 International Conference on Social Computing}}. \bibinfo{pages}{188--195}.
\newblock
\urldef\tempurl%
\url{https://doi.org/10.1109/SocialCom.2013.35}
\showDOI{\tempurl}


\bibitem[Vu(2021)]%
        {Vu2021-ASE}
\bibfield{author}{\bibinfo{person}{Duc-Ly Vu}.} \bibinfo{year}{2021}\natexlab{}.
\newblock \showarticletitle{py2src: Towards the Automatic (and Reliable) Identification of Sources for PyPI Package}. In \bibinfo{booktitle}{\emph{2021 36th IEEE/ACM International Conference on Automated Software Engineering (ASE)}}. \bibinfo{pages}{1394--1396}.
\newblock
\urldef\tempurl%
\url{https://doi.org/10.1109/ASE51524.2021.9678526}
\showDOI{\tempurl}


\bibitem[Vu et~al\mbox{.}(2021)]%
        {Vu2021-FSE}
\bibfield{author}{\bibinfo{person}{Duc-Ly Vu}, \bibinfo{person}{Fabio Massacci}, \bibinfo{person}{Ivan Pashchenko}, \bibinfo{person}{Henrik Plate}, {and} \bibinfo{person}{Antonino Sabetta}.} \bibinfo{year}{2021}\natexlab{}.
\newblock \showarticletitle{LastPyMile: Identifying the Discrepancy between Sources and Packages}. In \bibinfo{booktitle}{\emph{Proceedings of the 29th ACM Joint Meeting on European Software Engineering Conference and Symposium on the Foundations of Software Engineering}} (Athens, Greece) \emph{(\bibinfo{series}{ESEC/FSE 2021})}. \bibinfo{publisher}{Association for Computing Machinery}, \bibinfo{address}{New York, NY, USA}, \bibinfo{pages}{780–792}.
\newblock
\showISBNx{9781450385626}
\urldef\tempurl%
\url{https://doi.org/10.1145/3468264.3468592}
\showDOI{\tempurl}


\bibitem[Vu et~al\mbox{.}(2020)]%
        {Vu2020-EuroS}
\bibfield{author}{\bibinfo{person}{Duc-Ly Vu}, \bibinfo{person}{Ivan Pashchenko}, \bibinfo{person}{Fabio Massacci}, \bibinfo{person}{Henrik Plate}, {and} \bibinfo{person}{Antonino Sabetta}.} \bibinfo{year}{2020}\natexlab{}.
\newblock \showarticletitle{Typosquatting and Combosquatting Attacks on the Python Ecosystem}. In \bibinfo{booktitle}{\emph{2020 IEEE European Symposium on Security and Privacy Workshops (EuroS\&PW)}}. \bibinfo{pages}{509--514}.
\newblock
\urldef\tempurl%
\url{https://doi.org/10.1109/EuroSPW51379.2020.00074}
\showDOI{\tempurl}


\bibitem[Wang et~al\mbox{.}(2020)]%
        {Wang2020-ICSE}
\bibfield{author}{\bibinfo{person}{Ying Wang}, \bibinfo{person}{Ming Wen}, \bibinfo{person}{Yepang Liu}, \bibinfo{person}{Yibo Wang}, \bibinfo{person}{Zhenming Li}, \bibinfo{person}{Chao Wang}, \bibinfo{person}{Hai Yu}, \bibinfo{person}{Shing-Chi Cheung}, \bibinfo{person}{Chang Xu}, {and} \bibinfo{person}{Zhiliang Zhu}.} \bibinfo{year}{2020}\natexlab{}.
\newblock \showarticletitle{Watchman: Monitoring Dependency Conflicts for Python Library Ecosystem}. In \bibinfo{booktitle}{\emph{Proceedings of the ACM/IEEE 42nd International Conference on Software Engineering}} (Seoul, South Korea) \emph{(\bibinfo{series}{ICSE '20})}. \bibinfo{publisher}{Association for Computing Machinery}, \bibinfo{address}{New York, NY, USA}, \bibinfo{pages}{125–135}.
\newblock
\showISBNx{9781450371216}
\urldef\tempurl%
\url{https://doi.org/10.1145/3377811.3380426}
\showDOI{\tempurl}


\bibitem[Wu et~al\mbox{.}(2023)]%
        {Wu2023-SANER}
\bibfield{author}{\bibinfo{person}{Jianyu Wu}, \bibinfo{person}{Weiwei Xu}, \bibinfo{person}{Kai Gao}, \bibinfo{person}{Jingyue Li}, {and} \bibinfo{person}{Minghui Zhou}.} \bibinfo{year}{2023}\natexlab{}.
\newblock \showarticletitle{Characterize Software Release Notes of GitHub Projects: Structure, Writing Style, and Content}. In \bibinfo{booktitle}{\emph{2023 IEEE International Conference on Software Analysis, Evolution and Reengineering (SANER)}}. \bibinfo{pages}{473--484}.
\newblock
\urldef\tempurl%
\url{https://doi.org/10.1109/SANER56733.2023.00051}
\showDOI{\tempurl}


\bibitem[Wyss et~al\mbox{.}(2022)]%
        {Wyss2022-ICSE}
\bibfield{author}{\bibinfo{person}{Elizabeth Wyss}, \bibinfo{person}{Lorenzo De~Carli}, {and} \bibinfo{person}{Drew Davidson}.} \bibinfo{year}{2022}\natexlab{}.
\newblock \showarticletitle{What the Fork? Finding Hidden Code Clones in Npm}. In \bibinfo{booktitle}{\emph{Proceedings of the 44th International Conference on Software Engineering}} (Pittsburgh, Pennsylvania) \emph{(\bibinfo{series}{ICSE '22})}. \bibinfo{publisher}{Association for Computing Machinery}, \bibinfo{address}{New York, NY, USA}, \bibinfo{pages}{2415–2426}.
\newblock
\showISBNx{9781450392211}
\urldef\tempurl%
\url{https://doi.org/10.1145/3510003.3510168}
\showDOI{\tempurl}


\bibitem[Xiao et~al\mbox{.}(2022)]%
        {DBLP:conf/icse/XiaoHXTDZ22}
\bibfield{author}{\bibinfo{person}{Wenxin Xiao}, \bibinfo{person}{Hao He}, \bibinfo{person}{Weiwei Xu}, \bibinfo{person}{Xin Tan}, \bibinfo{person}{Jinhao Dong}, {and} \bibinfo{person}{Minghui Zhou}.} \bibinfo{year}{2022}\natexlab{}.
\newblock \showarticletitle{Recommending Good First Issues in GitHub {OSS} Projects}. In \bibinfo{booktitle}{\emph{44th {IEEE/ACM} 44th International Conference on Software Engineering, {ICSE} 2022, Pittsburgh, PA, USA, May 25-27, 2022}}. \bibinfo{publisher}{{ACM}}, \bibinfo{pages}{1830--1842}.
\newblock
\urldef\tempurl%
\url{https://doi.org/10.1145/3510003.3510196}
\showDOI{\tempurl}


\bibitem[Xu et~al\mbox{.}(2022)]%
        {Xu2022-FSE}
\bibfield{author}{\bibinfo{person}{Congying Xu}, \bibinfo{person}{Bihuan Chen}, \bibinfo{person}{Chenhao Lu}, \bibinfo{person}{Kaifeng Huang}, \bibinfo{person}{Xin Peng}, {and} \bibinfo{person}{Yang Liu}.} \bibinfo{year}{2022}\natexlab{}.
\newblock \showarticletitle{Tracking patches for open source software vulnerabilities}. In \bibinfo{booktitle}{\emph{Proceedings of the 30th ACM Joint European Software Engineering Conference and Symposium on the Foundations of Software Engineering}} (<conf-loc>, <city>Singapore</city>, <country>Singapore</country>, </conf-loc>) \emph{(\bibinfo{series}{ESEC/FSE 2022})}. \bibinfo{publisher}{Association for Computing Machinery}, \bibinfo{address}{New York, NY, USA}, \bibinfo{pages}{860–871}.
\newblock
\showISBNx{9781450394130}
\urldef\tempurl%
\url{https://doi.org/10.1145/3540250.3549125}
\showDOI{\tempurl}


\bibitem[Xu et~al\mbox{.}(2023)]%
        {Xu2023-ASE}
\bibfield{author}{\bibinfo{person}{Weiwei Xu}, \bibinfo{person}{Hao He}, \bibinfo{person}{Kai Gao}, {and} \bibinfo{person}{Minghui Zhou}.} \bibinfo{year}{2023}\natexlab{}.
\newblock \showarticletitle{Understanding and Remediating Open-Source License Incompatibilities in the PyPI Ecosystem}. In \bibinfo{booktitle}{\emph{Proceedings of the 38th IEEE/ACM International Conference on Automated Software Engineering}} (Kirchberg, Luxembourg) \emph{(\bibinfo{series}{ASE '23})}. \bibinfo{publisher}{Association for Computing Machinery}, \bibinfo{address}{New York, NY, USA}.
\newblock


\bibitem[Zhou and Mockus(2012)]%
        {Zhou2012-ICSE}
\bibfield{author}{\bibinfo{person}{Minghui Zhou} {and} \bibinfo{person}{Audris Mockus}.} \bibinfo{year}{2012}\natexlab{}.
\newblock \showarticletitle{What make long term contributors: Willingness and opportunity in OSS community}. In \bibinfo{booktitle}{\emph{2012 34th International Conference on Software Engineering (ICSE)}}. \bibinfo{pages}{518--528}.
\newblock
\urldef\tempurl%
\url{https://doi.org/10.1109/ICSE.2012.6227164}
\showDOI{\tempurl}


\bibitem[Zhou and Mockus(2015)]%
        {Zhou2015-TSE}
\bibfield{author}{\bibinfo{person}{Minghui Zhou} {and} \bibinfo{person}{Audris Mockus}.} \bibinfo{year}{2015}\natexlab{}.
\newblock \showarticletitle{Who Will Stay in the FLOSS Community? Modeling Participant’s Initial Behavior}.
\newblock \bibinfo{journal}{\emph{IEEE Transactions on Software Engineering}} \bibinfo{volume}{41}, \bibinfo{number}{1} (\bibinfo{year}{2015}), \bibinfo{pages}{82--99}.
\newblock
\urldef\tempurl%
\url{https://doi.org/10.1109/TSE.2014.2349496}
\showDOI{\tempurl}


\bibitem[Zhu et~al\mbox{.}(2016)]%
        {Zhu2016-FSE}
\bibfield{author}{\bibinfo{person}{Jiaxin Zhu}, \bibinfo{person}{Minghui Zhou}, {and} \bibinfo{person}{Audris Mockus}.} \bibinfo{year}{2016}\natexlab{}.
\newblock \showarticletitle{Effectiveness of Code Contribution: From Patch-Based to Pull-Request-Based Tools}. In \bibinfo{booktitle}{\emph{Proceedings of the 2016 24th ACM SIGSOFT International Symposium on Foundations of Software Engineering}} (Seattle, WA, USA) \emph{(\bibinfo{series}{FSE 2016})}. \bibinfo{publisher}{Association for Computing Machinery}, \bibinfo{address}{New York, NY, USA}, \bibinfo{pages}{871–882}.
\newblock
\showISBNx{9781450342186}
\urldef\tempurl%
\url{https://doi.org/10.1145/2950290.2950364}
\showDOI{\tempurl}


\bibitem[Zimmermann et~al\mbox{.}(2019)]%
        {Markus2019-USENIX}
\bibfield{author}{\bibinfo{person}{Markus Zimmermann}, \bibinfo{person}{Cristian-Alexandru Staicu}, \bibinfo{person}{Cam Tenny}, {and} \bibinfo{person}{Michael Pradel}.} \bibinfo{year}{2019}\natexlab{}.
\newblock \showarticletitle{Small World with High Risks: A Study of Security Threats in the npm Ecosystem}. In \bibinfo{booktitle}{\emph{28th USENIX Security Symposium (USENIX Security 19)}}. \bibinfo{publisher}{USENIX Association}, \bibinfo{address}{Santa Clara, CA}, \bibinfo{pages}{995--1010}.
\newblock
\showISBNx{978-1-939133-06-9}
\urldef\tempurl%
\url{https://www.usenix.org/conference/usenixsecurity19/presentation/zimmerman}
\showURL{%
\tempurl}


\end{thebibliography}

\end{document}